\newcommand{\rem}[1]{}
\newcommand{\comment}[1]{}

\newcommand{\beq}{\begin{equation}}
\newcommand{\eeq}{\end{equation}}

\newcommand{\beqa}{\begin{eqnarray}}
\newcommand{\eeqa}{\end{eqnarray}}

\newcommand{\refe}[1]{(\ref{#1})}

\newcommand{\bgeq}{\beq\left\{\begin{array}}
\newcommand{\egeq}{\end{array} \right.\eeq}

\newcommand{\pderc}[3]{\left.{\partial #1 \over \partial #2}\right|_{#3}}

\newcommand{\qav}[1]{\left\langle #1 \right\rangle}

\newcommand{\Gc}{{\cal G}}

\newcommand{\bgraf}{\left\{\begin{array}}
\newcommand{\egraf}{\end{array}\right.}

\newcommand{\bgeqn}{\[\left\{\begin{array}}
\newcommand{\egeqn}{\end{array} \right.\]}

\newcommand{\Nepo}{Nepomnyashchii}

\newcommand{\Av}{{\bf A}}
\newcommand{\xv}{{\bf x}}
\newcommand{\kv}{{\bf k}}
\newcommand{\qv}{{\bf q}}

\newcommand{\cho}{{\psi_{lo}}}
\newcommand{\alp}{\alpha}
\newcommand{\psit}{\tilde\psi}

\newcommand{\chiu}{\bar\psi}
\newcommand{\psib}{\bar\psi}

\newcommand{\Gcu}{\bar\Gc}
\newcommand{\Gb}{\Gamma}
\newcommand{\Gbb}{\bar\Gamma}

\newcommand{\disp}{\displaystyle}

\newcommand{\bO}{\beta\Omega}

\newcommand{\diag}[1]{}

\newcommand{\beql}[1]{\begin{equation}\label{#1}}

\newcommand{\boh}{}

\newcommand{\vd}{{\boh v}}

\newcommand{\ud}{{\boh u}}

\newcommand{\Check}[1]{}

\newcommand{\refE}[1]{Eq.~(\ref{#1})}

\newcommand{\rv}{{\bf r}}

\newcommand{\Pc}{{\cal P}}
\newcommand{\rhoo}{\rho}

\newcommand{\FORSE}[1]{}

\documentstyle[aps,multicol,prb,eqsecnum,psfig,feynmf]{revtex}

\begin{fmffile}{fmfRG}
\fmfcmd{%
style_def glt expr p =
 save dpp;
 numeric dpp;
 dpp = ceiling (pixlen (p, 10) / dash_len) / length p;
 for k=0 upto dpp*length(p)/2 -.5 :
  cdraw point k/dpp of p ..
   point (k+.3)/dpp of p;
 endfor
 cdraw subpath (length(p)/2, length(p)) of p;
enddef;
style_def gll expr p =
 save dpp;
 numeric dpp;
 dpp = ceiling (pixlen (p, 10) / dash_len) / length p;
 for k=0 upto dpp*length(p) -1 :
  cdraw point k/dpp of p ..
   point (k+.3)/dpp of p;
 endfor
enddef;
style_def gtt expr p =
 cdraw p;
enddef;
}

\def\narrowtext{\begin{multicols}{2} \global\columnwidth20.5pc}
\def\widetext{\end{multicols} \global\columnwidth42.5pc}
\input{epsf.tex}

\def\mytop#1{\vskip #1}
\def\mybottom#1{\vskip #1}

\begin{document}

\title{Renormalization Group Approach to the Infrared Behavior
of a Zero-Temperature Bose System}

\author{F.~Pistolesi$^1$, C.~Castellani$^2$, C.~Di Castro$^2$,
and G.C.~Strinati$^3$}

\address{
$^1$Laboratoire de Physique et Mod\'elisation des Milieux Condens\`es\\
CNRS-UJF,  B.P. 166 38042 Grenoble, France\\
$^2$Dipartimento di Fisica, Universit\`a ``La Sapienza'',
    UdR INFM, I-00185 Roma, Italy\\
$^3$Dipartimento di
    Fisica, Universit\`a di Camerino, UdR INFM,
    I-62032 Camerino, Italy
}

\date{\today}


\maketitle

\begin{abstract}
We exploit the renormalization-group approach to establish the {\em
exact} infrared behavior of an interacting Bose system at zero
temperature.
The local-gauge symmetry in the broken-symmetry phase is implemented
through the associated Ward identities, which reduce the number of
independent running couplings to a single one.
For this coupling the $\epsilon$-expansion can be controlled to all
orders in $\epsilon$ ($=3-d$).
For spatial dimensions $1 < d \leq 3$ the Bogoliubov fixed point is
unstable towards a different fixed point characterized by the
divergence of the longitudinal correlation function.
The Bogoliubov linear spectrum, however, is found to be independent
from the critical behavior of this correlation function, being exactly
constrained by Ward identities.
The new fixed point  properly gives a finite value of the
coupling among transverse fluctuations, but due to virtual intermediate
longitudinal fluctuations the effective coupling affecting the
transverse correlation function flows to zero.
As a result, no transverse anomalous dimension is present.
This treatment allows us to recover known results for the quantum Bose
gas in the context of a unifying framework and also to reveal the
non-trivial skeleton structure of its perturbation theory.
\end{abstract}

\narrowtext



\section{Introduction}
\label{SecI}

Perturbation theory for neutral zero temperature interacting bosons in the
superfluid phase has been developed in the fifties\cite{Bel,HP59}
following the mean-field result of Bogoliubov.\cite{Bog}
Notwithstanding the success of its immediate implementation in the low
density limit\cite{Bel} and in proving the coincidence of the
quasiparticle and sound modes,\cite{GN,HK,Jona,HM} the unresolved problem
of infrared (IR) divergences due to the Goldstone mode led to surprising
results.
As a matter of fact, much later A. \Nepo\ and Y. \Nepo\cite{Nepo}
(from here on referred as NN) demonstrated that the fundamental
quantity $\Sigma_{12}(k)$, the anomalous self-energy, must actually
vanish for vanishing momentum ($k\rightarrow 0$) and $d \leq 3$.
In the Bogoliubov approximation the existence of a linear spectrum
relies on the presence of $\Sigma_{12}(k)\neq 0$.
Since NN found that the linear spectrum is preserved for $k\rightarrow
0$, their result implies that the structure of the quadratic part of the action
must be actually much more complicated than the simple Bogoliubov form.

The NN results were obtained by studying directly the skeleton
structure of the diagrammatic perturbation theory (PT).
To determine the asymptotic behavior of the correlation functions
for $k\rightarrow 0$ NN had to
resort to a self-consistent analysis of the skeleton perturbation
theory, finding relations among set of diverging diagrams.
This method, besides being not fully
transparent, might not be readily controlled.
The need of developing a PT free from IR divergences was actually
realized by Popov\cite{Popov} before NN.
Popov development was achieved within a functional-integral approach,
by introducing a phase-amplitude representation for the bosonic
variables in the IR region for $|\kv|$ smaller than a cutoff
$k_o$.
In this way at all intermediate steps the calculation turned out to
be free from IR divergences.
However, elimination of the arbitrary parameter $k_o$ remains
nontrivial in Popov approach, as it requires full control of the
IR region up to momenta of the order of $k_o$ where the
phase-amplitude description becomes less and less accurate since at
large momenta it is the particle representation to be appropriate.
In addition, the phase-amplitude description somewhat obscures the
particle viewpoint even at small momenta. As a matter of fact, when
$k\rightarrow 0$ it is not easy to follow with the phase-amplitude description the
{\em nontrivial} evolution of the elementary excitations from free
particles to the sound mode.
Probably, it was for this reason that Popov recognized in his own
language the NN result about the vanishing of the anomalous
self-energy only at a later stage.\cite{PS}

The vanishing of the anomalous self-energy is not a mere mathematical
result but it has a definite {\em physical} origin.
In this particular context of broken gauge symmetry,
this finding reflects the general picture given by
Pata\v{s}inskij and Pokrovskij,\cite{PP} whereby the Goldstone-mode
singularity of the transverse correlation function drives a divergence
also in the longitudinal correlation function for a continuous broken
symmetry.
An additional interesting issue is to understand how,
technically, the IR divergences do not lead to a
anomalous dimension for the quadratic propagator, as it happens
instead for critical phenomena.

These circumstances lead to the necessity of providing a unifying
and fully controlled treatment of IR divergences.
In this respect, the renormalization group (RG) approach appears to be
the natural and reliable tool to determine the IR behavior of the
system in the presence of IR divergences.
In this paper, we exploit the RG approach to obtain the {\em exact\/}
IR behavior of the vertex functions (and thus of the correlation
functions) for a neutral system of interacting bosons in the
broken-symmetry phase at zero temperature.
To this end, we will make extensive use of the Ward identities (WI)
associated with  gauge invariance, which pose strong constraints on
the RG equations and enable us to obtain the desired solution to all
orders of the $\epsilon$-expansion (here $\epsilon=3-d$).
Although Popov method might at a first look be preferable to deal with the
broken-symmetry phase since it deals at the outset with
variables that are manifestly gauge invariant, it appears useful to
complement the phase-amplitude approach by studying the IR behavior of
the system via a more standard particle representation.
Our treatment allows us to follow the evolution of the propagator
continuously, from the particle region to the hydrodynamical one.
A short account of this work has been given in Ref.\ \onlinecite{nostro}.

Usually, the appearance of IR divergences is related to a second-order
phase transition, when a competition between two different phases with
equal free energy leads to divergent fluctuations at the critical
temperature.  In the present case, although the system is in a stable
phase and {\em no phase transition} occurs,
there exists a competition among degenerate states which are
associated with different values of the macroscopic phase of the order
parameter and are thus physically equivalent.
This leads to the divergence of the longitudinal correlation function
mentioned above, while all correlation functions obtained as averages
of local gauge-invariant operators (like the density-density
correlation function) are expected to be free from IR divergence.
A similar behavior is found for fermionic systems, where the perturbation
theory for the stable phase of the Luttinger liquid presents infrared
divergences that have to disappear when the symmetry is correctly
enforced through Ward identities.\cite{MetznerDicastro}

Perturbation theory for the broken symmetry phase in the particle
representation considers averages of operators which are not {\em
local} gauge invariant, and is therefore plagued by IR divergences at
intermediate steps of the calculation.
It is clear that these divergences cannot be independent from each
other as they have to cancel out when calculating averages of the
above-mentioned local gauge-invariant operators.
In this respect, Ward identities provide explicit connections among the
divergences.
In turn, this makes {\em all} local gauge-invariant susceptibilities
(like the compressibilities) finite and stabilizes the
system accordingly with respect to phase fluctuations.

By our RG approach we will be able to prove that the three available
gauge-invariant susceptibilities of the system (namely, the condensate
susceptibility, the ordinary susceptibility  related to the
sound velocity, and the total density) are indeed finite, being
invariants of the RG flow.
This will be done by establishing explicit connections among the
running couplings of the theory.
In this way, we will be left with only one independent running
coupling, whose infrared behavior will then be established to {\em
all} orders in $\epsilon$ ($=3-d$) for $d>1$.
The fixed point for $d<3$ is characterized by a finite interaction
coupling among the transverse fluctuating fields.
This at first sight suggests that the theory is interacting at the
fixed point like the standard $\phi^4$ theory.\cite{Amit}
We find, in fact, that this is not true due to the presence of the
longitudinal field.
One can define an effective interaction among transverse fluctuations
entering the relevant subset of diagrams responsible of the
appearance of the anomalous dimension in $\phi^4$ theory.
This effective interaction is flowing to zero at the
fixed point, thus preserving the $1/k^2$ behavior
for the transverse propagator.
The occurrence of the sound spectrum will turn out to be independent
from the scaling behavior of the independent running couplings, as it
depends only on the underlying gauge symmetry.
We will also recover the leading IR behavior of the single-particle
Green's functions and of the (two-particle) response functions obtained
previously by Gavoret and Nozi\`eres\cite{GN} and by NN\cite{Nepo}.

RG treatments of the zero-temperature interac\-ting-boson problem have
been previously given by Weich\-man\cite{Weich88} and by
Benfatto\cite{Benfatto}.  Weichman found the correct exponent of the
longitudinal susceptibility by performing a one-loop calculation for
an intrinsically non-divergent quantity (the free energy) and then
used a simple scaling argument to determine the singular behavior of
the longitudinal single-particle Green's function.
This scaling argument can, in general, be justified by the RG
approach, as it will be shown below by our treatment.
Benfatto, on the other hand, used a Wilson-like RG approach in a
rigorous fashion to determine the scaling behavior for a number of
running couplings.
However, his treatment was limited to $d=3$ and did not take advantage
of a systematic implementation of WI, which in turn would guarantee the linear
behavior of the spectrum irrespective of the IR behavior of the
longitudinal correlation function.
The fact that the theory is asymptotically free in $d=3$, as found by
Benfatto, is thus not essential to stabilize the superfluid phase by
preserving the linear spectrum.

The question naturally arises whether the momentum region, where
phase fluctuations (leading to IR divergences) dominate, has physical
relevance apart from establishing the superfluid behavior.
To this end, a generalized Ginzburg criterion can be introduced for
the momentum variable to determine, in particular, whether the IR
region will or will not merge into the Bogoliubov region,  where
the spectrum is still linear but the longitudinal correlation function
is approximately constant.
For the low-density Bose gas one can explicitly show\cite{PS} that the
Ginzburg region extends up to a characteristic momentum that vanishes
exponentially (in the gas parameter) with respect to the extension of
the linear Bogoliubov region.
In the more general case, it is not {\em a priori} possible to assess
whether the Bogoliubov region might be washed out by the growing of the
Ginzburg region.  In this case a full RG approach is required,
as shown in the present paper.

The plan of the paper is as follows.
In Section \ref{SecII} we recall how IR divergences manifest
themselves in PT by considering the one-loop correction to the
Bogoliubov approximation, and then set up the RG treatment
for the problem at hand.
In Section \ref{SecIII} we derive and
analyze the WI  relevant to our purposes.
In Section \ref{SecIV} we determine the exact IR behavior
of the vertex functions in the context of the
$\epsilon$ expansion.
Section \ref{SecVI} gives our conclusions.  For completeness,
additional technical points are discussed in the Appendices.
Specifically, Appendix \ref{AppCoupling} deals with the definition of
gauge invariance in terms of the couplings introduced in this paper,
Appendix \ref{App1loop} shows the one-loop evaluations of relevant
vertex functions and the validity of the WI,
Appendix \ref{AppRG1loop} gives the one-loop equations for the
running couplings entering the propagators.


\section{IR divergences and setup of the RG treatment}
\label{SecII}


In this Section, we briefly recall how the IR divergences appear in the
PT for an interacting boson system in the presence of a condensate
beyond the Bogoliubov approximation, and then set up the RG treatment
to heal these divergences.
To this end, we will exploit a functional-integral approach which is
especially suited to the purpose.\cite{Andersen}
%


\subsection{Bogoliubov approximation and the appearance of
IR divergences}
\label{SecIIA}


We consider a system of neutral bosons interacting through a
short-range potential $v$, in the presence of auxiliary external
sources $\lambda$, $\mu$, and $\Av$ that serve to generate the
field, density, and current connected correlation functions.
The associated action is given by:
\beqa
    S &=&
    \int \!\! dx \left\{ \vphantom{\int}
    \psi^*(x)
    \left[\partial_\tau+\mu(x)\right]
    \psi(x)
    -\left|(\nabla-i\Av)\psi(x)\right|^2
    \right.
    \nonumber \\
    &&\quad
    - {v\over 2}  \, |\psi(x)|^4
    \left. \vphantom{\int}
    +\psi(x)\lambda^*(x)+\psi^*(x)\lambda(x)
    \right\} \,.
    \label{r1}
\eeqa
In this expression, $x=(\tau,\rv)$ (with $\tau$ ranging from 0 to the
inverse temperature $\beta$) is a vector with $d+1$ dimensions,
$\psi(x)$ is a bosonic field obeying periodic boundary conditions over
the imaginary time $\tau$, and $v$ is the two-body interaction.
We shall always consider the zero temperature limit $\beta\rightarrow \infty$.
Since we are concerned with the IR behavior, we can choose $v$ local from
the outset, with an ultraviolet (UV) cutoff implied in UV divergent
integrals.
We have also set $\hbar=1$, $k_B=1$, and $m=1/2$.

The sources $\lambda(x)$ and $\Av(x)$ are let to vanish at
the end of the calculation, while $\mu(x)$ reduces to the ordinary
chemical potential $\mu$.
In the following, we shall sometimes use the short-hand
notation $A_{\nu} = (\mu,\Av)$ with $\nu=0,\dots,d$.

The relevant correlation functions can be obtained from the
free energy functional $F$, defined as:
\beq
    F[\lambda, \lambda^*, A_\nu]
    =
    \ln \int\!\!\! {\cal D}\psi^* {\cal D}\psi\,\,
        e^{S[\psi,\psi^*,\lambda,\lambda^*,A_\nu]}
    \, .
    \label{r5}
\eeq
In particular, the single-particle normal and anomalous
Green's functions are given by
\[
    \left\{
    \begin{array}{rcl}
    \Gc_{11}(x_1,x_2)
    &\equiv&
    -\qav{\psi(x_1)\psi^*(x_2)}_c
    = \displaystyle
    -{\delta F \over \delta \lambda^*(x_1) \delta \lambda(x_2)}
    \\ &&\\
    \Gc_{12}(x_1,x_2)
    &\equiv&
    -\qav{\psi(x_1)\psi(x_2)}_c
    = \displaystyle
    -{\delta F \over \delta \lambda^*(x_1) \delta\lambda^*(x_2)}
    \end{array}
    \right.
\]
where the connected average
$
\langle\psi^* \psi\rangle_c
=
\langle\psi^* \psi\rangle -
\langle\psi^* \rangle \langle\psi\rangle
$
is defined with respect to the weight function $\exp\{S\}$.

Bose condensation is conveniently introduced in terms of the Fourier
transforms
\beq
    \left\{
    \begin{array}{rcl}
    \psi(x)
    &=&
    (\bO)^{-1/2}\sum_{\omega_s, \kv}
    e^{i(\omega_s \tau+\kv \rv)} \psi(\omega_s, \kv)
    \\
    \psi^*(x)
    &=&
    (\bO)^{-1/2}\sum_{\omega_s, \kv}
    e^{-i(\omega_s \tau+\kv \rv)} \psi^*(\omega_s, \kv)
    \end{array}
    \right.
    \label{r7}
\eeq
where $\kv$ is a wave vector, $\omega_s=2\pi s \beta^{-1}$
(s integer) is a bosonic Matsubara frequency, and $\Omega$ is the
volume occupied by the system. We set
\beq
    \left\{
    \begin{array}{rcl}
    \psi(k) &=& \tilde\psi(k)+(\bO)^{1/2} \delta_{k,0} \alpha
    \\
    \psi^*(k) &=& \tilde\psi^*(k)+(\bO)^{1/2}\delta_{k,0} \alpha^*
    \end{array}
    \right.        ,
    \label{r11}
\eeq
where now $k=(\omega_s,\kv)$ and $\alpha$  is
arbitrary, and introduce \refE{r11} into the action \refe{r1}.
In this way, the action can be split into a (free) quadratic part
$S_Q=\sum_k \tilde\psi^*(k) \tilde \psi(k) \Gc_o^{-1}(k)$
with associated inverse propagator
\beq
    \Gc_o^{-1}(k) = i\omega_s -\kv^2+\mu
    \label{a17}
\eeq
and into the remaining perturbation terms.\cite{Bel}
The parameter $\alpha$ is then fixed by requiring that
$\langle\tilde\psi(k=0)\rangle = 0$,
which corresponds to the absence of ``tadpole'' diagrams,\cite{Popov}
{\em i.e.}, to the vanishing of 1-particle irreducible diagrams
with a single line entering:
\beq
    \parbox{20mm}{
    \begin{fmfgraph*}(20,10)
    \fmfleft{i} \fmfright{o}
    \fmf{fermion}{i,v} \fmf{phantom}{o,v}
    \fmfblob{.4w}{v}
    \end{fmfgraph*}
}
    =0 \,.
\eeq
The value of $\alpha$ thus becomes a function of the
uniform external source $\lambda$, and coincides with the (square root
of the) condensate density in the physical limit $\lambda \rightarrow 0$.
This requirement simplifies  PT considerably.
In particular, it is convenient to introduce a matrix
Green function by defining
$\psi_1(k)=\tilde \psi(k)$ and
$\psi_2(k)=\tilde \psi^*(-k)$.
In this way,
$
    \Gc_{11}(k) = \Gc_{22}^*(-k) = - \qav{\tilde\psi(k) \tilde\psi^*(k)}
$
and
$
    \Gc_{12} = \Gc_{21}^*(-k) = - \qav{\tilde\psi(k) \tilde\psi(-k)}
$
satisfy the Dyson-Beliaev equation
$
    {\Gc}(k)^{-1} = {G}^{-1}_o(k) - {\Sigma}(k)
$
with
\beq
    {G}_o =
    \left(
    \begin{array}{cc}
        \Gc_o(k) & 0 \\
        0 & \Gc_o(-k) \\
    \end{array}
    \right)
    \quad ,
    \label{r16}
\eeq
and the matrix elements of the self-energy $\Sigma$ are distinguished
by the number of incoming and outgoing lines.
The off-diagonal terms $\Sigma_{12}(k) = \Sigma^*_{21}(k)$
characterize the broken-symmetry state as they vanish identically in the
normal state.
Given $\Sigma(k)$, a simple matrix inversion leads to $\Gc$.
The Bogoliubov approximation consists in taking the diagrams with no
loops for the self-energy:
$ \Sigma_{11}(k) = 2 v |\alpha|^2$,
$ \Sigma_{12}(k) = \alpha^2 v $.
Within the same approximation, the vanishing of tadpole diagrams is
equivalent to the vanishing of the linear term in the action.
This gives for $\lambda=0$:
\beq
    v |\alpha|^2 = \mu
    \label{r21}
\eeq
and the following form for the Green's functions:
\beq
    \Gc_{11}(k)
    =
    -{i\omega_s+\kv^2+\mu
    \over
    \omega_s^2+ E_\kv^2 }
    \qquad
    \Gc_{12}(k)
    =
    {\mu \over \omega_s^2+E_\kv^2 }
    \label{r22}
\eeq
with $E_{\kv}^2 = \kv^4+2\mu \kv^2$.
Note that, within the Bogoliubov approximation, the occurrence of the
sound mode ({\em i.e.}, $E_{\kv} \propto |\kv|$ for $\kv^2 \ll 2 \mu$)
is strictly related to $\Sigma_{12}(k=0)$ being non-vanishing.

Inclusion of higher-order terms beyond the Bogoliubov approximation,
however, spoils the above results due to the occurrence of IR
divergences.
For instance, let us consider the following one-loop diagram contributing
to $\Sigma_{12}$:
\beq
    A(k) =
    \parbox{20mm}{\fmfframe(0,3)(0,3){
    \begin{fmfgraph*}(20,10)
    \fmfleft{i} \fmfright{o}
    \fmf{fermion,label=$-k$}{i,v1} \fmf{fermion,label=$k$}{o,v2}
    \fmf{fermion,right,tension=.4,label=$q$}{v1,v2}
    \fmf{fermion,right,tension=.4,label=$k+q$}{v2,v1}
    \end{fmfgraph*}}}
    \sim
    \alpha^2 \, v^2\, c_o \int^\Lambda\!\!\!
    {d^{d+1} \tilde q \over \tilde q^2(\tilde k+\tilde q)^2}
    \label{r24}
\eeq
where $\Lambda$ is an UV cutoff and $\tilde q$ is a
$(d+1)$-dimensional vector.
We have used the small $k$-form of the Bogoliubov propagators
\refe{r22} that behave behave like $\tilde k ^{-2}$, where $\tilde k =
(\omega_s/c_o, \kv)$ with $c_o^2 = 2 \mu$ being the square of the
sound velocity.
Problems arise because $A(k)$ diverges when $d \leq 3$ as $k
\rightarrow 0$ (in the zero-temperature limit) as follows:
\beq
    A(k) \sim
    \left\{
    \begin{array}{ll}
        k^{d-3} & {\rm for}\ d<3 \\
        \ln(k^2/\Lambda^2) & {\rm for}\ d=3
    \end{array}
    \right. \, ,
    \label{r25}
\eeq
while it remains finite for $d>3$.
The one-loop contribution \refe{r24} thus overcomes the Bogoliubov
contribution for small enough $k$.
It is then natural to introduce in this context a generalized Ginzburg
criterion for the momentum variable, by comparing one-loop and
zero-loop self-energies:
\beq
    1 \approx {\Sigma_{12}^{(1)}(k_G) \over \Sigma_{12}^{(0)}(k_G)}
    = (v^3 n)^{1/2} \,
    \left\{ \begin{array}{ll}
     k_G^{d-3}  & d<3 \\
    \log(c_o/k_G) & d=3\,.
    \end{array}
    \right.
    \label{rnew1}
\eeq
Here, we have used $\alpha^2 \approx n$ and we
set the UV cut-off $ \Lambda$ equal to $c_o$ as simple estimates.
According to this criterion, IR divergences become dangerous when $k
\lesssim k_G$, where $ k_G = (v^3 n)^{1\over 2(3-d)}$ for $d<3$ and
$k_G = c_o \exp\{-(v^3 n)^{-1/2}\}$ for $d=3$ as given in
Ref.~\onlinecite{PS}.

Notwithstanding the presence of these  singularities,
explicit calculations have shown that IR divergences
cancel out in most physical quantities.\cite{Bel}
A notable example is the Hugenholtz-Pines identity\cite{HP59},
namely,
\beq
    \Sigma_{11}(k=0) - \Sigma_{12}(k=0) = \mu
    \label{r26}
\eeq
which implies cancellation of the IR divergences in the two self-energies.

The identity \refe{r26} provides the simplest example
of WI, whereby the (irreducible) diagrams of PT
are constrained by the underlying gauge symmetry.
A systematic study of the implications of WI was originally performed
in Ref. \cite{GN}.
In the following Sections we will exploit extensively the whole set of
WI to reduce the set of IR divergent diagrams of PT to only a few
independent ones.

As mentioned in the Introduction, not all IR divergences,
however,  disappear from the theory. In particular, the longitudinal
single-particle Green's function (to be discussed in detail in Section
\ref{SecIV}) has to diverge in the IR on physical grounds,\cite{PP,Fisher}
 implying that
\beq
    \Sigma_{12}(k=0) = 0
    \label{r27}
\eeq
for $d \leq 3$.
This was found as an {\em exact } result by NN.
It is clear that the result \refe{r27} which depends on
dimensionality cannot be inferred from a WI like
\refE{r26}, which is instead independent from dimensionality.
It is also clear that WI by themselves are not sufficient to
specify completely the IR behavior of the system.
To this purpose,
specific dynamical properties have to be taken into account in
addition to WI, as it will be discussed in Section \ref{SecIV} where
the RG equation for a single independent running coupling will be
derived.


\subsection{Representation
for the Field-Theory RG treatment of interacting bosons}
\label{SecIIB}


The RG treatment in the broken-symmetry phase is considerably
simplified when replacing the Bose field $\psi$ and $\psi^*$ by their
longitudinal $(\psi_l)$ and transverse ($\psi_t$) components to the
direction of the broken symmetry:\cite{Benfatto}
\beq
    \left\{
    \begin{array}{rcl}
    \psi(x)  &=& \alpha + \tilde\psi_l(x) + i \tilde \psi_t(x)
    \\
    \psi^*(x)  &=& \alpha + \tilde \psi_l(x) - i \tilde\psi_t(x)
    \end{array}
    \right.
    \label{r28}
\eeq
where $\tilde\psi_l$ and $\tilde\psi_t$ are real functions,
and the order parameter has been taken real as well
without loss of generality.
These new fields are introduced to distinguish explicitly between
transverse and longitudinal fluctuations, whose difference is not
evidenced by the standard representation resulting into \refE{r22}.

\widetext
\mytop{5mm}

In terms of Fourier transform of the new fields, the action \refe{r1}
becomes for $\bf A=0$ and $\mu(x)=\mu$:
\beqa
    \lefteqn{S[\psit_i] =
    \bO
    \left[
        \mu \alp^2   -{v\over 2} \alp^4 +\lambda_l\alpha
    \right]
    +(\bO)^{1/2}
    \left[
        2\mu  \alp -2v\alp^3 +\lambda_l
    \right] \psit_l(0)
    + (\bO)^{1/2} \lambda_t \psit_t(0)
}
    \nonumber \\
    &&
    +\sum_k
    \left\{
        \left[  \mu -3v\alp^2-\kv^2\right]
            \psit_l(k)\psit_l(-k)
        +\left[ \mu  -v\alp^2 -\kv^2\right]
            \psit_t(k)\psit_t(-k)
        - 2\omega_\nu
        \psit_l(-k) \psit_t(k)
    \right\}
    \nonumber \\
    &&
    -2v \alp (\bO)^{-1/2}\!\!\!\!\!\!\!\!\!\!\sum_{k_1+k_2+k_3=0}
    \left\{
        \psit_l(k_1) \psit_l(k_2) \psit_l(k_3)
        +\psit_l(k_1) \psit_t(k_2) \psit_t(k_3)
    \right\}
    \nonumber \\
    &&
    -(\bO)^{-1}\!\!\!\!\!\!\!\!\!\!\!\sum_{k_1+k_2+k_3+k_4=0}
    \!\!\!\!\!\!\!{v \over 2}
    \left\{
         \psit_l(k_1) \psit_l(k_2) \psit_l(k_3) \psit_l(k_4)
        +\psit_t(k_1) \psit_t(k_2) \psit_t(k_3) \psit_t(k_4)
        +2\psit_l(k_1)\psit_l(k_2)\psit_t(k_3)\psit_t(k_4)
    \right\} \,.
    \label{r33}
\eeqa

\mybottom{5mm}
\narrowtext

Quite generally, the order parameter is given by
\beq
    \psi_{io}(x) \equiv \langle \psi_i(x)\rangle =
    {\delta F /\delta \lambda_i(x)}
    \label{r35}
\eeq
where $i=(l,t)$, $\lambda=(\lambda_l+i \lambda_t)/2$, and $F$ is given
by \refE{r5} (once expressed in terms of $\lambda_l$ and $\lambda_t$).
Consistently with our choice,
we shall eventually assume the limit $\lambda_t=0$ and
$\lambda_l\rightarrow 0$, so that only $\psi_{lo}$ survives.

The variable $\alpha$ can be eliminated by the tadpole condition.
It can be useful to clarify this point in more details here,
since at higher orders elimination of $\alpha$ becomes more subtle.
Given the action \refe{r33}, one can regard the quadratic part as the
free theory and treat the remainder as a perturbation, including
the linear term in $\psit_l$ (we set $\lambda_i=0$ from the outset).
The diagrammatic perturbation theory is then well defined.
Diagrams are conveniently summed up by grouping them according
to the number of loops present in each of them.
One can verify that this is equivalent to scale the action as
$S\rightarrow S/a$ and calculate all quantities by an expansion in the
parameter $a$.
A naive way of proceeding would be to choose the value of $\alpha$
such that the linear term in the action vanishes,
giving  $\alpha^2\,v=\mu$.
In this way, propagators are massless and all physical quantities
can be calculated within the loop-expansion
with the parameter $\alpha$ fixed to its mean-field value
$\sqrt{\mu/v}$.
Actually, the presence of the 3-legs vertices makes this approach
cumbersome.
As a matter of fact, all diagrams will have tadpole corrections.
It is not difficult to recognize that the effect of these corrections
could be incorporated in a redefinition of $\alpha$.
The shift required would be exactly the correction to the
average of $\psi_l$ induced by the tadpole corrections:
$\qav{\psi_l} = \alpha + \langle \psit_l \rangle  \neq \alpha $.
An {\em alternative} procedure is thus to fix $\alpha$ by the
condition that tadpole 1-particle irreducible diagrams vanish.
This condition guarantees that $\qav{\psi_l} = \alpha $,
since the vanishing of 1-particle irreducible diagrams implies also
the vanishing of all reducible ones contributing to $\qav{\psi_l}$.
The equation for $\alpha$ is then calculated perturbatively
as a loop-expansion without tadpole insertions:
\beq
     F^{(0)}(\alpha)+F^{(1)}(\alpha)+F^{(2)}(\alpha)+\dots = 0
     \label{condensa}
\eeq
where $F^{(0)}(\alpha)=2\alpha(\mu-v \alpha^2)$ and  $F^{(n)}(\alpha)$ is
the sum of all one-leg n-loop diagrams (cfr.\ \refe{AB1} and
\refe{AB2} for the one-loop case).
Substituting the loop expansion of $\alpha$
($\alpha=\alpha^{(0)}+\alpha^{(1)}+\dots $) and solving
for $\alpha$ order by order, one obtains
$\alpha^{(0)}=\mu/v$,
$\alpha^{(1)}= -F^{(1)}(\alpha^{(0)})/
(\partial F^{(0)}(\alpha)/\partial \alpha)_{\alpha=\alpha^{(0)}}
$, $\dots$ .
This procedure defines a much simpler diagrammatics, since
no tadpole diagrams are present, their contribution
appearing as a shift of $\alpha$.
The price to pay is that one has to calculate all diagrams for
arbitrary $\alpha$  both in the 3-leg couplings and
propagators.
This implies that the propagators would have a mass (gap),
but since all quantities have to be expanded around
their mean-field values one recovers massless propagators in all
expressions.\cite{BraatenNieto}
This fixes the procedure in a precise way.

In our case an important simplification occurs.
As we are interested in the IR behavior of the exact propagators,
finite corrections are not important.
Since the tadpole diagrams
are all finite in the infrared, we will not calculate the finite
corrections to $\alpha$ accordingly.
In the following, $\alpha$ can be then thought as
fixed and to coincide with $\psi_{lo}$.
At the same time, the propagators are massless and no
tadpole diagrams will be considered.
In practice, this means that finite corrections to the condensate
are not important down to $d=1$. At that point
the condensate vanishes due to IR divergences.
We will see that precisely at $d=1$ our approach breaks down since the
the longitudinal and transverse fluctuations becomes equally singular.

As far as the chemical potential entering the action \refe{r33}
is concerned, it is possible to leave its value unspecified
in the following calculation since the interacting boson system
with a fixed chemical potential is well defined.\cite{mufixed}

PT simplifies by introducing the Legendre
transform of $F$ with respect to the sources $\lambda$:
\beq
    \Gamma[\psi_{io}, A_\nu] =
    \int \!\! dx\,  \lambda_i(x) \psi_{oi}(x) -
    F[\lambda_i[\psi_{oi}], A_\nu]
    \label{r36}
\eeq
where
\beq
    {\delta \Gamma /\delta \psi_{io}(x)} = \lambda_i(x)
    \label{r36a}
\eeq
and
\beq
    {\delta \Gamma / \delta A_{\nu}(x)} =
    - {\delta F / \delta A_\nu(x)} \, .
    \label{r36c}
\eeq
Similarly, the connected correlation functions and
the vertex functions are respectively given by:
\beqa
    {\delta^{(n+m)} F
    \over
    \delta \lambda_{i_1}(x_1) \dots \delta \lambda_{i_n}(x_n)
    \delta A_{\nu_1}(y_1) \dots \delta A_{\nu_m}(y_m)
    }
    &=& \nonumber\\
    -\Gc_{i_1\dots i_n;\nu_1\dots \nu_m}
    (x_1, \dots, x_n; y_1, \dots, y_m) \,, &&
    \label{r37}
\eeqa
\beqa
    {\delta^{(n+m)} \Gamma
    \over
    \delta \psi_{i_1 o}(x_1) \dots \delta \psi_{i_n o}(x_n)
    \delta A_{\nu_1}(y_1) \dots \delta A_{\nu_m}(y_m)  }
     &=&
    \nonumber \\
    \Gamma_{i_1\dots i_n;\nu_1\dots \nu_m}
    (x_1, \dots, x_n; y_1, \dots, y_m) \, . &&
    \label{r38}
\eeqa
Here, the derivatives with respect to $\lambda_i$ give the connected
and amputated (1-particle irreducible)
correlation functions, for $F$ and $\Gamma$, respectively
The derivatives with respect to $A_\nu$ generate the current and
density response functions (or their amputated counterparts).

The single-particle Green's functions analogous to \refE{r22}
can be obtained from the quadratic part of the action \refe{r33},
which reads in terms of the independent fields:
\beq
    S_Q = \sum_{k,i,j}
    \psit^*_i(k) M_{ij}(k) \psit_j(k)
    \label{r39}
\eeq
with
\beq
    M(k) = \left(
        \begin{array}{cc}
            \mu-3v \alp^2-\kv^2 & -\omega_s \\
            \omega_s & \mu-v \alp^2 -\kv^2
        \end{array}
        \right)
    \label{r40} \, .
\eeq
With the aid of the condition \refe{r21}, one then obtains:
\beq
    \Gc(k)=  {1\over 2} (M^{-1})^t =
    -{1\over 2}
    {   1
        \over
        \omega_s^2+E_\kv^2
    }
    \left(
    \begin{array}{cc}
        \kv^2      & \omega_s \\
        -\omega_s  & 2\mu+\kv^2
    \end{array}
    \right)
    \label{r41}
\eeq
which for small $\kv$ and $\omega_s$ behaves like
\beq
    \left\{
    \begin{array}{rcl}
    \Gc_{tt}(k)
        &\sim& -c_o^2/(\omega_s^2+ c_o^2 \kv^2)\\
    \Gc_{lt}(k)
        &\sim& -\omega_s/(\omega_s^2+ c_o^2 \kv^2)\\
    \Gc_{ll}(k)
        &\sim& -\kv^2/(\omega_s^2+ c_o^2 \kv^2)
        \quad.
    \end{array}
    \right.
    \label{r42}
\eeq
In the present case the low-energy region is
dominated by the Goldstone (sound) mode.
The IR limit of the correlation functions is thus naturally defined
by taking frequency and wave vector to vanish while maintaining
their ratio  {\em constant}. For example, one can take
$
\lim_{\gamma \rightarrow 0} \Gc_{ij}(\gamma \omega_s, \gamma \kv)
$.
In this way, the three propagators \refe{r42} have {\em different} IR
behavior, with the strongest singularity residing in the transverse
propagator $\Gc_{tt}$.
This situation has to be contrasted with the standard
$\psi$-representation, where the Bogoliubov propagators
\refe{r22} share instead the {\em same} IR behavior.
The choice of $(\psi_l, \psi_t)$ in the place of $(\psi,\psi^*)$ will
thus turn out to be crucial to select the interaction terms on the
basis of their relevance.
Furthermore, this choice will be important from a physical point of
view since an anomalous dimension will appear in the longitudinal
component $\Gc_{ll}$ {\em only}.

To treat these singularities, it is important to classify the
coupling terms in the action \refe{r33} according to their
relevance in generating IR divergences.
A standard way to perform this task is to define scaled fields and
frequencies, so that the infrared dimensions of the propagators in
\refe{r42} coincide with their engineering dimensions.
In our case the scaling is defined as follows:
\beq
    \left\{ \begin{array}{rcl}
        \bar\psi_l(k_0, \kv) &=&
        c_o \tilde \psi_l(c_o k_0, \kv)\\
        \bar\psi_t(k_0, \kv) &=&
        \tilde \psi_t(c_o k_0, \kv)
    \end{array}
    \right.
    \label{r43}
\eeq
with the short-hand notation $k_0 = i \omega_s/c_o$
(from now on $k=(k_0,\kv)$).
Since the rescaled fields have dimensions $ [\chiu_l(k) ] =
[\psit_l(k) ] +1 = 0 $ and $ [\chiu_t(k)] = [\psit_t(k)] = -1$ (we
have defined the scale of the momentum as the unit, $[k]=1$), this leads
to the desired dimensions for the propagators of the rescaled fields:
\[
    [\Gcu_{tt}(k)] = -2\,,
    \qquad
    [\Gcu_{lt}(k)] = -1\,,
    \qquad
    [\Gcu_{ll}(k)] = 0\,.
\]
Since different diagrams contributing to the same quantity must have
the same dimension, couplings of higher dimension
must appear together with most divergent Green's functions.
Thus the engineering dimensions of the couplings for the new
fields can be used to classify their relevance.
Equation \refe{r43} implies that, in real space, the fields are
rescaled as
\beq
    \left\{ \begin{array}{rcl} \chiu_l(x_0, \xv) &=& c_o^{1/2}
    \chiu_l(x_0/c_o, \xv)\\ \chiu_t(x_0, \xv) &=&
    c_o^{-1/2}\chiu_t(x_0/c_o, \xv)\,, \end{array} \right.
    \label{r46}
\eeq
which follows from a redefinition of the Fourier transform:
\beq
    \chiu_i(k_0,\kv) =
    {1 \over \sqrt{\beta c_o \Omega}}
    \int_0^{\beta c_o}\!\!\!\!\! d x_0
    \!\int\!\! d^d \xv\, e ^{ -i k x} \chiu_i(x)
    \label{r48}
    \, .
\eeq
The dimension of the fields are $[\psi_l(x)]=(d+1)/2$ and
$[\psi_t(x)]=(d-1)/2$.
From the above considerations, the dimension
of a generic coupling $v_{n_l n_t}$ associated with the monomial
$\chiu_l^{n_l}(x) \chiu_t^{n_t}(x)$ is given by
\beq
    [v_{n_l,n_t} ]
    =
    -n_l {d+1\over 2} -n_t {d-1 \over 2}+d+1
    \,.
    \label{r50}
\eeq
For later convenience, we write explicitly the
dimensions of the couplings  up to four legs:
\beqa
    \bgraf{rcl}
        \left[v_{l}\right] &=& 2-\epsilon/2\\
        \left[v_{t} \right] &=& 3-\epsilon/2
    \egraf,
    \bgraf{rcl}
        \left[v_{ll} \right] &=& 0\\
        \left[v_{lt} \right] &=& 1\\
        \left[v_{tt} \right] &=& 2
    \egraf,
&& \label{r51a}
\\
    \bgraf{rcl}
        \left[v_{lll} \right] &=& -2+{\epsilon \over 2}\\
        \left[v_{llt} \right] &=& -1+{\epsilon \over 2}\\
        \left[v_{ltt} \right] &=& {\epsilon \over 2}\\
        \left[v_{ttt} \right] &=& 1+{\epsilon \over 2}
    \egraf,
    \bgraf{rcl}
        \left[v_{llll} \right] &=& -4+\epsilon\\
        \left[v_{lllt} \right] &=& -3+\epsilon\\
        \left[v_{lltt} \right] &=& -2+\epsilon\\
        \left[v_{lttt} \right] &=& -1+\epsilon\\
        \left[v_{tttt} \right] &=& \epsilon
    \egraf
&&
    \label{r51}
\eeqa
where $v_l=v_{10}$, $v_{ll}=v_{20}$, $v_{ltt}=v_{12}$, etc., and
$\epsilon=3-d$ parameterizes the distance from the
critical dimension $d_c=3$ at which logarithmic divergences appear.

Couplings with more than four legs have not been considered
since they are irrelevant at $d=d_c$, {\em i.e.}, they have negative
dimensions.
The vertex functions have the same dimensions of the
relative coupling constants. This implies, for instance,
that the tadpole equation is non-divergent in the infrared,
as anticipated, since $[v_l]=[\Gamma_l]=2-\epsilon/2>0$

To make PT calculations (as well as the
RG treatment) more convenient, we generalize the original
action \refe{r33} in the broken-symmetry phase as follows.
Besides the terms already present in \refe{r33}, we
introduce  additional running
couplings which are relevant or marginal at $d_c$
(while all irrelevant couplings were already present
in the original action).
The generalized action thus acquires the form:\cite{foot1}
\widetext
\beqa
    \lefteqn{S[\psib_i] =
    \bO v_{0}
    -(\bO)^{1/2} v_{l} \psib_l(0)
    -(\bO)^{1/2} v_{t} \psib_t(0)
        }\nonumber \\
    &&
    -{1\over 2!} \sum_k
    \left\{ \vphantom{k^2}
        [v_{ll} +z_{ll}\kv^2]\psib_l(-k)\psib_l(k)
        +2[v_{lt}+w_{lt} \omega_s] \psib_l(-k)\psib_t(k)
                \vphantom{k^2}
        +[v_{tt} +u_{tt} \omega_s^2 +z_{tt} \kv^2] \psib_t(-k)\psib_t(k)
    \right\}
    \nonumber
    \\ &&
    - {(\bO)^{-1/2}\over 3!} \sum_{k_1+k_2+k_3=0}
    \left\{
         v_{ttt} \psib_t(k_1)\psib_t(k_2)\psib_t(k_3)
        +3 v_{ltt} \psib_l(k_1)\psib_t(k_2)\psib_t(k_3)
        +v_{lll} \psib_l(k_1)\psib_l(k_2)\psib_l(k_3)
    \right\}
    \nonumber \\ &&
    - {(\bO)^{-1} \over 4!} \!\!\!\! \sum_{\sum_i k_i=0}
    \!\!\!\!\!\!
    \left\{
        v_{tttt}
        \psib_t(k_1)\psib_t(k_2)\psib_t(k_3)\psib_t(k_4)
        +6v_{lltt}
        \psib_l(k_1)\psib_l(k_2)\psib_t(k_3)\psib_t(k_4)
        +v_{llll}
        \psib_l(k_1)\psib_l(k_2)\psib_l(k_3)\psib_l(k_4)
    \right\} \,.
    \nonumber \\
    && \label{r53}
\eeqa
\narrowtext
The reason to use these new couplings in the place of the original ones
is related to the fact that they appears as coefficients
of the monomials of fields, and thus have a simple bare
scaling behavior. It is also clear that there is a definite
relation among the new set of couplings and the original ones,
that can be inferred by direct comparison of
\refe{r33} and \refe{r53}.
Since the new set of couplings is larger, it is also clear that they
will not be independent when related to the original parameters.
In Section \ref{SecIIIa} these connections will be discussed
and related to the gauge invariance of the original theory.

As far as the marginal couplings ($w_{lt}$, $u_{tt}$, $z_{tt}$) are
concerned, we recall that in the original action \refe{r33} they had,
respectively, the values (2,0,2) regardless of the original
parameters ($v,\mu,\alpha$).
We follow here a standard procedure of the field-theoretical RG
treatment and let the coefficients of the marginal terms proportional
to $\omega$, $\omega^2$, and $\kv^2$  assume arbitrary values in
order to renormalize the theory.
In principle, an alternative procedure would be to introducing two
wave-function-like renormalization parameters and one dynamical
renormalization parameter which, in turn, allows one to keep the matrix of
the propagators unchanged.
It will turn out, however, that the renormalization of these
parameters, although nontrivial, will not affect the scaling form of
the matrix of the renormalized propagators, since these parameters will
appear there only in some specific combinations that are independent
of the RG flow.
It will, in fact, turn out that the anomalous behavior of the
longitudinal propagator depends only on the renormalization of the coupling
$v_{ll}$; at the same time, the dynamical index will
stay finite at $z=1$, consistently with the Bogoliubov sound mode.

Quite generally, the matrix entering the quadratic part of the action
\refe{r53} is defined as in \refe{r39} and is given by:
\beq
    {M} =
    -\left(\begin{array}{cc}
        v_{ll} + z_{ll}\kv^2, & v_{lt} + w_{lt} \omega_s \\
        v_{tl} - w_{lt}\omega_s, & v_{tt} + u_{tt}\omega_s^2
            +z_{tt} \kv^2
    \end{array}
    \right)
    \label{r55}
\eeq
in the place of \refE{r40}.
The corresponding single-particle propagator becomes:
\beqa
    \bar\Gc(k)
    &=&
    -\left(\begin{array}{cc}
        v_{tt} +u_{tt} \omega_s^2 + z_{tt} \kv^2 \,,&
             -v_{tl}+ w_{lt} \omega_s \\
        -v_{lt}- w_{lt} \omega_s \,,& v_{ll}+z_{ll} \kv^2
        \end{array}
    \right)
    {1\over D(k)}
    \nonumber \\
    &\equiv&
    \left(\begin{array}{cc}
        \bar\Gc_{ll}(k) & \bar\Gc_{lt}(k)\\
        \bar\Gc_{tl}(k) & \bar\Gc_{tt}(k)
        \end{array}
    \right)
    \label{r57}
\eeqa
with
\beqa
    D(k) &=&
     v_{ll}v_{tt}-v_{lt}^2
    +(u_{tt} v_{ll} + w_{lt}^2)\omega_s^2
    \nonumber \\
    &&
    +(v_{ll} z_{tt} +v_{tt} z_{ll} + z_{ll} z_{tt}\kv^2) \kv^2
    \label{r58}
\eeqa
in the place of \refE{r41}.
For later convenience, the components of the propagator \refe{r57}
are depicted schematically as follows:
\beq
    \Gc_{ll}(k) =
    \parbox{10mm}{
    \begin{fmfgraph*}(10,3)
     \fmfleft{i}\fmfright{o}\fmf{gll}{i,o}
    \end{fmfgraph*}}\,,\
    \Gc_{lt}(k) =
    \parbox{10mm}{
    \begin{fmfgraph*}(10,3)
     \fmfleft{i}\fmfright{o}\fmf{glt}{i,o}
    \end{fmfgraph*}}\,,\
    \Gc_{tt}(k) =
    \parbox{10mm}{
    \begin{fmfgraph*}(10,3)
     \fmfleft{i}\fmfright{o}\fmf{gtt}{i,o}
    \end{fmfgraph*}}
    \,.
    \label{ar59}
\eeq
The remaining terms of the action beyond the quadratic part
are considered as perturbation.
The interaction terms retained in the RG treatment are:
\beq
    v_{ltt} =
    \parbox{8mm}{
    \begin{fmfgraph*}(8,12)
    \fmftop{t}\fmfright{r}\fmfleft{l}
    \fmf{gll}{t,v}\fmf{gtt}{r,v,l}\fmfdot{v}
    \end{fmfgraph*}}
    \,,\
    v_{tttt}=
    \parbox{12mm}{
    \begin{fmfgraph*}(10,8)
    \fmfleft{l1,l2}\fmfright{r1,r2}
    \fmf{gtt}{r1,v,r2}\fmf{gtt}{l1,v,l2}
    \fmfdot{v}
    \end{fmfgraph*}}
    \label{ar60}
\eeq
PT treatment then proceeds now along the usual lines.

Although the generalized action \refe{r53} contains four marginal
couplings ($v_{ll}$, $w_{lt}$, $u_{tt}$, $z_{tt}$) and seven relevant
couplings ($v_l$, $v_t$, $v_{lt}$, $v_{tt}$, $v_{ltt}$, $v_{ttt}$,
$v_{tttt}$), it will be shown in the next Section via the use of WI
that these couplings are actually not independent from each other.

\section{Gauge invariance and Ward Identities}
\label{SecIII}

The original action \refe{r1} satisfies local-gauge
invariance, in the sense that it is unaffected
by the following transformation:
\bgeq{lcllcl}
    \psi(x) &\rightarrow & \psi(x) e^{i\chi(x)}
    &
    \psi(x)^* &\rightarrow& \psi(x)^*e^{-i \chi(x)}\\
    \lambda(x) &\rightarrow & \lambda(x) e^{-i\chi(x)}
    &
    \lambda(x)^* &\rightarrow& \lambda(x)^*e^{i \chi(x)}\\
    \mu(x) &\rightarrow& \mu(x) - i\partial_\tau \chi(x)
    &
    \Av(x) &\rightarrow& \Av(x) - {\bf \nabla} \chi(x)
    \label{r60} \, .
\egeq
This invariance provides definite connections (in the form of WI)
among the vertex functions \refe{r38}.\cite{HK,Jona,HM}
These identities have been exploited by Gavoret and
Nozi\`eres to establish  connections
between vertex functions and thermodynamic derivatives for
vanishing wave vector and frequency.
In this Section, we shall make use of the WI
in two additional ways.
On the one hand, we shall {\em constraint}
the various couplings of the generalized action \refe{r53}
by showing that they are not independent from each other when
the generalized action is required to be gauge
invariant.
On the other hand, we will exploit the WI to
connect the IR singular parts of the vertex functions in PT
and to implement their cancellation to all orders in
the response functions.

To be consistent with our notation, we first
give a short derivation of the set of WI relevant to our purposes.
In terms of the longitudinal and transverse components
($\psi_l$, $\psi_t$), the original action \refe{r1} has the
following invariance:
\beq
    S[R_{ij} \psi_j, R_{ij}\lambda_j, A_\nu - \partial_\nu \chi]
    =
    S[\psi_i, \lambda_i, A_\nu]
    \label{r61}
\eeq
where $i$ and $j$ take the values $(l,t)$
and a summation over repeated indices is implied.
In \refE{r61} we have introduced the notation
$\partial_\nu = \left( i \partial_\tau, \nabla\right)$ and
$\bf R$ is the rotation matrix
\beq
    {\bf R} =
    \left(
        \begin{array}{cc}
        \cos \chi(x)& -\sin \chi(x) \\

        \sin \chi(x)& \cos \chi(x)
        \end{array}
    \right) \, ,
    \label{r62}
\eeq
$\chi(x)$ being an arbitrary real function.
The above invariance leads, in turn, to the following
invariance of the functional $\Gamma$:
\beq
    \Gamma[A_\nu-\partial_\nu \chi(x), R_{ij} \psi_{jo}]
    =
    \Gamma[A_\nu,\psi_{io}]
    \, .
    \label{r63}
\eeq
The desired connections among the vertex functions
are obtained at this point by differentiating \refE{r63} with
respect to $\chi$. One obtains:
\beq
    \Gamma_i(x) \sigma_{ij} \psi_{jo}(x)
    + \partial_\nu \Gamma_{;\nu}(x)
    = 0
    \label{r64}
\eeq
where $\sigma$ is the infinitesimal generator of rotations
\beq
    \sigma =
    \left(  \begin{array}{cc}
            0& -1\\ 1 & 0
        \end{array}
    \right)
    \nonumber
\eeq
and $\chi$ has been set to zero eventually.
Successive differentiations with respect to $\psi_{io}$ give in
addition:
\widetext
 \beq
    \Gamma_{im}(x_1,x_2) \sigma_{ij} \psi_{jo}(x_1)
    + \Gamma_i(x_1) \sigma_{im} \delta(x_1-x_2)
    + \partial^{x_1}_\nu \Gamma_{m;\nu}(x_2;x_1) = 0 \,,
    \label{r67}
\eeq
\beqa
    &&\Gamma_{imn}(x_1,x_2,x_3) \sigma_{ij} \psi_{jo}(x_1)
    +\Gamma_{im}(x_1,x_2) \sigma_{in}\delta(x_1-x_3)
    + \Gamma_{in}(x_1,x_3) \sigma_{im} \delta(x_1-x_2)
    \nonumber \\
    &&
    + \partial^{x_1}_\nu \Gamma_{mn;\nu}(x_2,x_3;x_1) = 0\, ,
    \label{r68}
\eeqa
and
\beqa
    &&
    \Gamma_{imnr}(x_1,x_2,x_3,x_4) \sigma_{ij} \psi_{jo}(x_1)
    +\Gamma_{imn}(x_1,x_2,x_3) \sigma_{ir} \delta(x_1-x_4)
    +\Gamma_{imr}(x_1,x_2,x_4) \sigma_{in}\delta(x_1-x_3)
    \nonumber \\
    &&
    + \Gamma_{inr}(x_1,x_3,x_4) \sigma_{im}\delta(x_1-x_2)
    + \partial^{x_1}_\nu \Gamma_{mnr;\nu}(x_2,x_3,x_4;x_1) = 0\, .
    \label{r69}
\eeqa
\narrowtext
The WI \refe{r64}-\refe{r69} are sufficient for the
RG treatment to be discussed in Section \ref{SecIV}.

For homogeneous external sources,
in terms of  Fourier transforms (for instance,
$
    \Gamma_i(x) = {1\over \beta \Omega} \sum_q e^{iqx} \Gamma_i(q)
$
with the short-hand notation $qx=\omega_s \tau+\kv \cdot \rv$)
the WI \refe{r64}-\refe{r69} simplify considerably.
In particular, the sets of
WI which will be explicitly used in the following
for vanishing transverse source (i.e., $\psi_{to}=0$)
can be grouped as follows:
\beq
    \Gb_t \cho  = 0\,,
    \label{r75}
\eeq
\beqa
    \Gb_{tl}(k)  \cho
    + \Gb_t  - i k_\nu \Gb_{l;\nu}(-k)
    &=& 0
    \label{r76} \\
    \Gb_{tt}(k)  \cho
    - \Gb_l
    - i k_\nu \Gb_{t;\nu}(-k)
    &=& 0 \,
    \label{r77}
\eeqa
from \refe{r64} and \refe{r67}, which encompass the Hugenholtz-Pines\cite{HP59}
identity for $k=0$;
\beqa
    &&
    \Gb_{ttl}(k_1,k_2) \cho
    +\Gb_{tt}(-k_2)
    -\Gb_{ll}(k_1+k_2)
    \nonumber \\
    &&
    - i (k_1)_\nu \Gb_{tl;\nu}(k_2,-k_1-k_2) = 0
    \, ,
    \label{r79}
\eeqa
\beqa
    &&
    \Gb_{ttt}(k_1,k_2) \cho
    -\Gb_{lt}(-k_2)
    -\Gb_{lt}(k_1+k_2)
    \nonumber \\
    &&
    - i (k_1)_\nu \Gb_{tt;\nu}(k_2,-k_1-k_2) = 0
    \,
    \label{r80}
\eeqa
which follow from \refe{r68} and are the standard WI associated with
the continuity equation modified by the presence of the 3-legs
vertices;
and from \refe{r69} the four legs ones
\beqa
    &&
    \lefteqn{
    \Gb_{tttt}(k_1,k_2,k_3)\cho
        -\Gb_{ltt}(-k_2-k_3,k_2)
    }
    \nonumber \\    &&
        -\Gb_{ltt}(k_1+k_3,k_2)
        -\Gb_{ltt}(k_1+k_2,k_3)
    \nonumber \\    &&
    - i(k_1)_\nu \Gb_{ttt;\nu}(k_2,k_3,-k_1-k_2-k_3) = 0
    \, .
    \nonumber \\
    \label{r81}
\eeqa
Here, $k_\nu = (i \omega_s, \kv)$ with $\nu=(0,\dots,d)$ and
the conservation of momentum has been used to eliminate the
dependence on the last momentum of each vertex function,
 {\em i.e.}, $\Gamma_i(k) = \Gamma_i(k=0) \equiv \Gamma_i$,
$\Gamma_{ij}(k_1,k_2) \equiv \Gamma_{ij}(k_1)$, and so on.
It is thus understood that the vertex functions are free from the
overall momentum conserving delta function.
We include the $i$ in the definition of $k_0$ only in the scalar
product entering the WI to simplify the notation; everywhere else
$k_0=\omega_s$.
The rescaling \refe{r43} has not been considered explicitly in the
present Section, but it can be readily introduced whenever necessary.

We recall that the WI \refe{r75}-\refe{r81} have been obtained using
the invariance properties of the original action \refe{r33}.
For arbitrary values of the couplings, the generalized action
\refe{r53} will not be invariant under the transformation \refe{r61}.
Thus the requirement of fulfilling \refe{r61} will provide constraints
on the couplings appearing in \refe{r53}.
It is also clear that this requirement by itself is not sufficient to
reduce the number of independent couplings to the original ones.
It is, in fact, possible to add to the original action \refe{r1} terms
of the type $ v_n |\psi|^{2n} $ still preserving gauge
invariance.  To identify \refe{r33} exactly with \refe{r53} one
should specify, for instance, that no 5-leg couplings are present.
With this additional condition, it can be shown that the
independent couplings left are only four (for instance,
$v_l$, $v_{ll}$, $\psi_{lo}$, and $\psi_{to}$)
corresponding to
$\mu$, $v$, $\lambda_l$, and $\lambda_t$ of the original action \refe{r1}.


In Appendix \ref{AppCoupling}, we discuss in more details
the formal definition of gauge invariance in terms of the couplings
introduced in this Section.
There we show how to enforce the invariance \refe{r60}
directly on the action.
In an equivalent way here we
proceed by assuming that the new action \refe{r53} shares
the same invariance properties of \refe{r1}.
Then \refE{r63} holds and the above WI follow.
We thus {\em enforce} the vertex functions associated with the new
action to satisfy these WI and obtain relationships among the bare
couplings of the action \refe{r33}.

To be more precise, we can calculate all vertex functions by a
loop expansion, since the WI are preserved order by order.
We can use this method to:
({\em i}) Constraint the {\em bare} couplings using the
lowest order 0-loops expressions;
({\em ii}) Connect the divergent
contributions to different {\em renormalized} running couplings
at any loop and thus find relations among themselves.


\subsection{Global Ward identities: $k=0$}
\label{SecIIIa}


We are now in a position to discuss the vanishing of the relevant bare
couplings $v_{l}$, $v_t$, $v_{lt}$, $v_{tt}$, and $v_{ttt}$
present in \refe{r53} at any dimension.

At the lowest order in the loop expansion, Eqs.~\refe{r75}-\refe{r77}
and \refe{r80} can be evaluated at vanishing $k$ and the vertex
functions expressed in terms of the bare couplings.
This procedure provides definite relationships among the
relevant bare couplings.
Specifically, we obtain $v_t=0$ from \refE{r75}  for non vanishing
condensate density $\psi_{lo}$;
$v_{lt} \psi_{lo} + v_t= 0$ from \refE{r76};
$v_{tt} \psi_{lo}-v_l = 0$ from \refE{r77};
and $v_{ttt}\psi_{lo}-2v_{lt} = 0$ from \refE{r80}.
This gives:
\beq
    v_{t}=0\,, \qquad v_{lt} =0\,, \qquad v_{ttt} =0\,,
    \label{r81a}
\eeq
and
\beq
    v_{l} = \psi_{lo} v_{tt}
    \label{r81b}
\eeq
for vanishing $\psi_{to}$.
To determine $v_l$, we consider in addition the ``equation of state''
\refe{r36a} (with $i=l$) for an isolated system with $\lambda_i=0$.
This is the counterpart of the condensate equation
\refe{condensa} in the new variables and fixes the
bare coupling $v_l=0$.
Actually, this is not true at higher orders, since the
shift of the condensate will induce finite corrections
to $v_l$ that must be fixed in order to let tadpole
diagrams vanish.
Since we are not interested in finite corrections, we
can set the bare and renormalized couplings $v_l=0$,
and thus $v_{tt}=0$.
The vanishing of $v_{tt}$ implements the Hugenholtz-Pines identity,
leading to gapless propagators.

We pass now to discuss the connections among the marginal couplings.
Let's consider the {\em bare} marginal couplings first.  The
identities \refe{r79} and \refe{r81} can be interpreted at the
lowest-order of the loop expansion as follows:
\beqa
    v_{ltt} \psi_{lo} - v_{ll} &=& 0
    \label{r87}
    \\
    v_{tttt} \psi_{lo} - 3 v_{ltt} &=& 0
    \label{r88}
\eeqa
where all $k$ have been taken to vanish.
The marginal bare couplings thus eventually
reduce to four independent ones $v_{ll}$, $w_{lt}$, $u_{tt}$,
and $z_{tt}$.

It is clear that the relations \refe{r81a}, \refe{r81b}, \refe{r87},
and \refe{r88} must be enforced on the bare running couplings
to guarantee the gauge invariance of the theory.
A field theoretic RG treatment is now possible only
if we neglect all irrelevant running couplings.
Although this spoils gauge invariance, it does not
matter for what concerns the leading order-singularities.
We can verify that gauge invariance is lost if we set all irrelevant
running couplings to zero, since in this case the WI (at the
lowest-order in loops) are no longer compatible with finite values of the
marginal running couplings.
In other words, the use of a perturbation theory for the
$\Gamma$'s with only the marginal running couplings will violate
the WI (as shown explicitly in Appendix \ref{App1loop}),
but the leading diverging contribution to each
$\Gamma$ function is fully retained by this approximation.
This implies that the finite part of the WI can be
violated, but the divergent contributions coming from different
$\Gamma$ functions will correctly simplify in the WI.
Thus the WI hold for the renormalization of each
vertex function, since these are introduced
exactly to subtract out the divergent parts.
For this reason the WI \refe{r87}-\refe{r88}
will hold for the {\em renormalized} running couplings too.


\subsection{Local Ward identities: small $k$}
\label{localWI}


There remains to consider the connections between the couplings
($w_{lt}$, $u_{tt}$, $z_{tt}$), associated with the wave vector and
frequency dependence in the action \refe{r53}, with the thermodynamic
derivatives in the limit of vanishing $\kv$ and $\omega$.
To this end, we consider the vertex functions $\Gamma_{ll}$,
$\Gamma_{lt}$, and $\Gamma_{tt}$ associated with the couplings
of interest.
The long-wavelength limit of these vertex functions
has been obtained long ago.\cite{GN,HK,Jona}
Here, we will rederive them in a more compact form
taking into account the presence of IR divergences.
Important information on the scaling of the running coupling
will follow from these connections.

We begin by establishing their symmetry properties.
From previous considerations
$\Gamma_{lt}(k=0)=0$ and $\Gamma_{tt}(k=0)=0$.
Using the symmetry properties of the action \refe{r53} under
space and time reversal and under exchange of the two components
$l$ and $t$ of the bosonic field, one can show that
$\Gamma_{lt}(k)$ is odd under the replacement
$\omega_s \rightarrow -\omega_s$ while $\Gamma_{tt}$ is even.
We then take the zero-temperature limit
(whereby $\omega_s$ becomes a continuous variable, denoted by $\omega$
from now on) and consider the WI \refe{r76} and \refe{r77}.

It is convenient to cast the vertex functions
in a form that makes their symmetry properties manifest.
For instance, from
$\Gamma_{lt}(\omega,\kv) = -\Gamma_{lt}(-\omega,\kv)$
we write
\beq
    \Gamma_{lt}(\omega,\kv) = \omega\, \Pc_{lt}(\omega,\kv^2)
    \label{r89}
\eeq
where $\Pc_{lt}(\omega,\kv^2)$ is an even function of $\omega_s$.
Similarly, we can write
\beq
    \Gamma_{l;0}(k) = \Pc_{l;0}(\omega,\kv^2)
    \label{r90}
\eeq
and
\beq
    \Gamma_{l;n}(k) = \omega\, i k_n\, \Pc_{l;v}(\omega,\kv^2)
    \label{r91}
\eeq
($n=1,\dots,d$) where $\Pc_{l;0}(k)$ and $\Pc_{l;v}(k)$
are even functions of $\omega$ and depends on $\kv$ only through $\kv^2$
(note that the suffix $v$ is here a reminder that
we are considering a vector function).
In this way, the WI \refe{r76} becomes:
\beq
    \psi_{lo} \Pc_{lt}(k) + \Pc_{l;0}(k)
    + \kv^2 \Pc_{l;v}(k) = 0
    \label{r92}
\eeq
where \refE{r75} has been taken into account.

To the purpose of understanding the behavior of the
singularities when we let $\omega$ or $\kv$  vanish indepentently,
we discuss the form of the divergences.
At the critical dimension ($d_c=3$), a generic vertex function
$\Gamma(k)$ of a single external variable in the limit $k\rightarrow
0$ can be written formally as
\beqa
    \Gamma(k) &=&
    (k^2)^{D/2}
    \left[
    \sum_{m=1}^{\infty}
    A_m \ln^m\left(\kv^2+{\omega^2\over c_o^2}\right)
    + B\right]
    \nonumber \\
    && +\,  C
    \label{r93}
\eeqa
where $D$ stands for the IR dimension of $\Gamma(k)$
and $A_m$, $B$, and $C$ are constants.
[In general, instead of the term $(k^2)^{D/2}$ one can
use $(\omega^2)^{D/2}$, $(\kv^2)^{D/2}$, or any linear combination
of the two.]
In this expression, we are not considering the terms arising from the
internally diverging diagrams, since the divergent contributions
due to these diagrams are eliminated by the renormalization
at lower order of the other $\Gamma$ functions.
In this scheme, the divergences left are all due to
the primitively divergent diagrams of each vertex function.
These divergences must cancel out in the WI, so
exact relations result among the primitively divergent contributions
to each $\Gamma$ function.
The form \refe{r93} implies also that the $k\rightarrow 0$ limit of the
divergent $\log$ series does not depend on which limit is taken first,
$\omega\rightarrow 0$ or $\kv\rightarrow 0$.
This is a  consequence
of the fact that all singular dependence comes through the
single argument of the $\log$ terms. This fact can be verified
order by order in loops [see Appendix \ref{App1loop}].
Of course, the order in which the limits are taken affects the final value
of the $\Gamma$ functions that have overall $\omega$ and $\kv$ terms.
The $\Gamma$ functions that have $D=0$ will be totally symmetric,
and the information gained for one of the two variables set
to zero applies also for $k\rightarrow 0$.
Knowing the degree of divergence of the vertex functions and the
explicit combinations of $\omega$ and $\kv$ appearing in each
expression, one can establish relations between the vertex functions
in presence of IR divergences.

In order to proceed further we need the IR dimension ({\em i.e.}, the
engineering dimension after the rescaling \refe{r43}) of the composite
vertex parts.
One should consider, however, that IR dimension is correctly defined
for monomials of fields.
For instance, we can readily obtain that the dimension of
$\Gamma_{l;l^2}$ is $2-{\epsilon/2}$, where by $l^2$ we mean the
field $\psi_l^2$ ($\Gc_{;l^2}(x) \equiv -\qav{\psi_l^2(x)}$).
But the dimension of a physical composite fields will
mix different monomials and thus different IR behavior:
$\Gamma_{l;0}=\Gamma_{l;l^2}+\Gamma_{l;t^2}$.
As we are concerned with divergences, we identify the IR dimension of
a composite field with the most divergent dimension (in the above
example with $[\Gamma_{l;t^2}]=-{\epsilon/2}$ we define
$[\Gamma_{l;0}]=-{\epsilon/2}$).

We report the IR dimension of other relevant
$\Gamma$ functions for further reference:
\beq
    \bgraf{rcl}
        \left[\Gbb_{l;j} \right]   &=& 2-\epsilon\\
        \left[\Gbb_{t;0} \right]   &=& 1-\epsilon\\
        \left[\Gbb_{t;j} \right]   &=& 2-\epsilon   \,, \\
    \egraf\quad
    \bgraf{rcl}
        \left[\Gbb_{;00} \right]   &=& -\epsilon\\
        \left[\Gbb_{;i0} \right]   &=&  2-\epsilon\\
        \left[\Gbb_{;ij} \right]   &=& 4-\epsilon
            \,.
    \egraf
    \label{r51b}
\eeq
These identities are obtained by performing the Fourier transform of
the Green's function, taking into account that square of fields
depend on a single variable.\cite{Amit}

We pass now to discuss the WI \refe{r92}.
Since the functions entering \refe{r92} have vanishing IR
dimensions [cf.~Eqs.~\refe{r51b} and \refe{r51} taking also into accont
that the functions $\Pc$ differ from the respective
$\Gamma$ by factors $\omega$ or $\kv$],
the last term on the left-hand side of \refe{r93}
vanishes as $\kv^2$ $\ln^m(\kv^2+\omega^2/c_o^2$ in the IR limit.
In this way we obtain:
\beq
    \psi_{lo} \lim_{k\rightarrow 0} \Pc_{lt}(k) =
    -\lim_{k\rightarrow 0} \Pc_{l;0}(k)
    \label{r94}
\eeq
(which is related to $w_{lt}$)
where
\beq
    \lim_{k\rightarrow 0}
    \Pc_{l;0}(k) = {1\over \beta\Omega}
    {\partial \Gamma \over \partial \psi_{lo} \partial \mu}
    \label{r95}
\eeq
is a thermodynamic derivative by definition.
We can also obtain the same
results by setting $\kv=0$ and letting $\omega\rightarrow 0$.
The final result is valid for $k \rightarrow 0$, since
$\Pc_{l;0}(k)$  is a function
of $\kv^2+\omega^2/c_o^2$ only for $k\rightarrow 0$.

We consider next the WI \refe{r77}.
Expressing again the vertex functions $\Gamma$ in terms of the
$\Pc$ functions,
$\Gamma_{t;0}=\omega \Pc_{t;0}$
and
$\Gamma_{t;i}=i k_i \Pc_{t,v}$
 as in Eqs.~\refe{r89}-\refe{r91}, we write:
\beq
    \psi_{lo} \Gamma_{tt}(k)+\omega^2 \Pc_{t;0}(k)
    + \kv^2 \Pc_{t;v}(k) = 0
    \label{r96}
\eeq
from which we obtain for $\omega=0$
\beq
    \lim_{\kv \rightarrow 0}
    {\Gamma_{tt}(0,\kv)\over \kv^2}
    =
    -{1\over \psi_{lo}}
    \lim_{\kv\rightarrow 0}
    \Pc_{t;v}(0,\kv^2)
    \label{r97}
\eeq
(related to $z_{tt}$), and for $\kv={\bf 0}$
\beq
    \lim_{\omega \rightarrow 0}
    {\Gamma_{tt}(\omega,{\bf 0})\over \omega^2}
    =
    -{1\over \psi_{lo}}
    \lim_{\omega\rightarrow 0}
    \Pc_{t;0}(\omega,{\bf 0})
    \label{r98}
\eeq
(related to $u_{tt}$).
Note that the function $\Pc_{t;v}$ has dimension $1-\epsilon/2$.
This suggests that the primitively divergent contribution to this function
vanishes at $d_c$.
By inspection of the PT one can verify that this is actually the case
since the current insertions reduce the
strength of the divergence.
This fact is important since it implies the absence of
singularities on the left-hand side of \refE{r97}
(and thus the fact that $z_{tt}$ will not scale).
This does not imply that $\Pc_{t;v}$ vanishes in the limit of
$k\rightarrow0$, since the non-singular contribution can be finite.

The function $\Pc_{t;\nu}(k)$ in the limit $k\rightarrow 0$
cannot be related directly to thermodynamic derivatives,
since these involve derivatives with respect to the transverse
field $\psi_{to}$.
In order to identify the two limits
\refe{r97} and \refe{r98}, we use the following additional
WI (obtained by differentiating Eq.~\refe{r64} with respect to
$A_\mu$ and then taking the Fourier transform):
\beq
    \psi_{lo} \Gamma_{t;\mu}(k) +
    i k_\nu \Gamma_{;\nu \mu}(k) = 0
    \label{r99}
\eeq
which in terms of the $\Pc$ functions reads (with a factor $\omega$
divided out)
\beq
    \psi_{lo} \Pc_{t;0}(k)-
    \Gamma_{;00}(k) -
    \kv^2 \Pc_{;v0}(k) = 0
    \label{s1}
\eeq
and
\beqa
    &&i k_n \psi_{lo} \Pc_{t;v}(k)-i \omega^2 k_n \Pc_{;0v}(k)
    \nonumber \\
&&
    +i k_m\left[
    {k_m k_n \over \kv^2} \phi_l(k)
    +\left(\delta_{m,n} - {k_m k_n \over \kv^2} \right)
    \phi_t(k)
    \right]
    \label{s2}
\eeqa
$(m,n=1,\dots,d)$, where a sum of repeated indices is implied
and the quantity within brackets is $\Gamma_{;mn}$.
Equation \refe{s1} for $\kv=0$ gives
\beq
    \psi_{lo} \lim_{\omega\rightarrow 0} \Pc_{t;0}(k)
    =  \lim_{\omega \rightarrow 0} \Gamma_{;00}(k)
    = {1\over \beta \Omega}
    {\partial^2 \Gamma \over \partial \mu^2}
    \label{s3}
    \,,
\eeq
which is again a thermodynamic derivative and relates
$u_{tt}$ to $\Gamma_{;00}$.
Multiplying \refE{s2} by $-ik_n$ and summing over $n$ we obtain further
\beq
    \psi_{lo} \Pc_{t;v}(k)-\omega^2 \Pc_{0;0v}(k)
    +\phi_l(k) = 0\,,
    \label{s4}
\eeq
which for $\omega=0$ and $\kv\rightarrow 0$ yields
\beq
    \psi_{lo} \lim_{\kv\rightarrow 0} \Pc_{t;v}(k)
    = -\lim_{\kv\rightarrow 0} \phi_l(k)
    = -{n\over m} \,.
    \label{s5}
\eeq
Here, $m(=1/2)$ is the mass and the last identity follows from
the Galilean invariance of the action.\cite{GN}
Equation \refe{s5} relates $z_{tt}$ to $n/m$.
Finally, in the limit $k\rightarrow 0$ the limit of
$\Gamma_{ll}(k)$ is
$(\beta \Omega)^{-1} \partial^2\Gamma/\partial \psi_{lo}^2$
by definition.

We have thus verified that, even in presence of
IR divergences, the asymptotic limit of the quadratic
vertex functions and of its derivatives are related to
thermodynamic derivatives.
Summarizing, in the IR limit one expects that
$\Gamma_{lt}(k) = - \Gamma_{l;0}(0)\, \omega/\psi_{lo} $ and
$\Gamma_{tt}(k) =  (2\,n\, \kv^2 +  \Gamma_{;00}(0)\,\omega^2)/\psi_{lo}^2 $.

Using the scaling of the RG obtained in the
next Section, we obtain further that the quantities
\beqa
\Gamma_{ll}(\rho k)&\sim& v_{ll}(\rho)\, , \qquad\qquad
\Gamma_{lt}(\rho k) \sim \omega \rho \, w_{lt}(\rho) \, ,
\nonumber\\
\Gamma_{tt}(\rho k) &\sim& \rho^2
\left[
   \omega^2 \,u_{tt}(\rho) + \kv^2 \, z_{tt}(\rho)
\right]
\label{smallrho}
\eeqa
when the RG scale $\rho$ (defined in the next section)
goes to zero.

These connections between the $k\rightarrow0$ limits
of the vertex functions and the asymptotic limits of the
RG couplings allow us to identify the asymptotic
$\rho\rightarrow0$ limit of the ratio
\beq
    {c^2(\rho)\over c_o^2} =
    {v_{ll}(\rho) z_{tt}(\rho)
    \over
    v_{ll}(\rho) u_{tt}(\rho)+w_{lt}(\rho)^2
    }
    \label{s6}
\eeq
with the macroscopic sound velocity $c_s$. To this end, we write
\beqa
    \lim_{\rho\rightarrow0}
    c^2(\rho)
    &=&
    {
    {\displaystyle \partial^2 \Gamma
    \over
    \displaystyle \partial \psi_{lo}^2}
        \displaystyle
     {1\over \psi_{lo}^2} {n\over m}
    \over
    \displaystyle
    {\partial^2 \Gamma \over  \partial \psi_{lo}^2}
    \left(  -{1\over\psi_{lo}^2}
        {\partial^2 \Gamma\over \partial \mu^2}
    \right)
    + {1\over \psi_{lo}^2}
    \left(
    {\partial^2\Gamma
    \over
    \partial \psi_{lo} \partial \mu}
    \right)^2
    }
    \nonumber \\
    &=&
    {n/m \over (dn/d\mu)_\lambda} \,,
    \label{s7}
\eeqa
which equals the square of the {\em macroscopic} sound velocity
by standard thermodynamic arguments.
This last identification is obtained by expressing derivatives
at constant $\psi_{lo}$ in terms of physical derivatives at constant
external sources $\lambda$.
The ratio \refe{s6} enters the determinant \refe{r58},
where it identifies the (square of the) {\em microscopic}
sound velocity.
The constant term ($v_{ll} v_{tt}-v_{lt}^2$) vanishes identically as
explained above.

Finally, we note in this context that the ratio
\beq
    c_o^2 \lim_{\rho\rightarrow0}
    {v_{ll}(\rho) \over w_{lt}(\rho)}
    =
    {
    {\displaystyle \partial^2 \Gamma \over
    \displaystyle \partial \psi_{lo}^2}
    \over
    {\displaystyle  1\over \displaystyle \psi_{lo}}
    {\displaystyle \partial^2 \Gamma \over
    \displaystyle \partial \psi_{lo}\partial \mu}
    }
    =
    -2\,\psi_{lo}^2
    \left.{d \mu \over \displaystyle d \psi_{lo}^2}\right|_\lambda
    \label{s8}
\eeq
is proportional to (the inverse of) the ``condensate''
compressibility taken at constant external sources,
which makes it a truly physical quantity.
We thus expect on physical grounds that the ordinary
compressibility $d n /d\mu$
and the ``condensate'' compressibility
$d n_o / d \mu$  (as well as
the density $n$) are all {\em finite}, since the physical
system we are considering is in a stable phase.
As a consequence, singular PT contributions will have to
cancel out exactly in these particular combinations of
running couplings, so that the ratios \refe{s6} and
\refe{s8} as well as $z_{tt}(\rho)$ (which would give an anomalous
dimension to the transverse field) will not change
under the RG flow.\cite{MetznerDicastro}
In Section \ref{SecIV} we will present a diagrammatic
proof of this fact for the two couplings $u_{tt}$ and
$w_{lt}$, while the argument given after Eq.~\refe{r98} is
valid to all orders in PT.


\section{Field-theoretic RG treatment}
\label{SecIV}


In this Section, we set up and solve the RG
equations for the flow of the coupling constants discussed in
the previous Sections. We will explicitly calculate the
one-loop PT contributions to the relevant vertex functions,
and show that higher orders of PT do not modify the IR behavior
found at the one-loop level.
In other words, we will be able to assess the {\em exact}
IR behavior of the interacting Bose gas at zero temperature.
This result contrasts with the standard (e.g., $\phi^4$) theory
of critical phenomena, whereby critical exponents can
be calculated only approximately.
The physical origin of this simple behavior in our case stems on the
fact that the same theory is free from IR divergences when formulated
in the variables phase and amplitude.\cite{phasefree}
In our variables the theory remains interacting at the fixed point
to give a diverging longitudinal susceptibility.
Nevertheless, we will show that the
effective interaction entering the transverse
vertex part $\Gamma_{tt}$ vanishes exactly.
This explains the difference with the $\phi^4$ theory in the final
result for the $\Gc_{tt}$ correlation functions and
recovers in an effective way the asymptotic freedom for the
phase-amplitude action.

As usual, we shall discard all  irrelevant running couplings of
\refe{r53} and write equations for the four marginal couplings
$(v_{ll}, w_{lt}, u_{tt}, z_{tt})$.
Before considering our specific problem, we
summarize the RG procedure employed shortly.\cite{DombGreen,Amit}


\subsection{RG procedure}
\label{RGApproach}


Given the set of bare running couplings $\{g^o_i\}$ of the
theory (regularized in the UV with a cutoff $\Lambda$), we
impose the following normalization conditions:
\beql{rr49}
    \Gamma_{i}(k)|_{k^2=\kappa^2} = g_i \,.
\eeq
The renormalized vertices are defined in the following way:
\beql{rr50}
    \Gamma_R(k_j,g_i(g^o_i,\kappa,\Lambda), \kappa)
    =
    \Gamma(k_j,g^o_i,\Lambda)\,.
\eeq
Differentiating Eq.~\refe{rr50} with respect to $\kappa$
at  constant $g^o_i$ we have:
\beq
    \left[
    \kappa \left.{\partial \over \partial \kappa}\right|_{g}
    + \kappa \left.{\partial g_i\over \partial \kappa}\right|_{g^o}
    {\partial \over \partial g_i}
    \right]
    \Gamma_R(k_i, g, \kappa)
    =0
    \,.
    \label{rr51}
\eeq
If the dimension of the coupling $g_i$ is $\alpha_i$,
we can introduce the dimensionless couplings $u_i$ and $u^o_i$:
$ g_i = \kappa^{\alpha_i} u_i$ and
$g^o_i = \kappa^{\alpha_i} u_i^o$.
The RG equation takes then its standard form
\beql{rr56}
    \left[
        \kappa\pderc{}{\kappa}{u}
        + \beta_i(u) \pderc{}{u_i}{\kappa}
    \right] \Gamma_R(k_j,u_i,\kappa)
    = 0\,,
\eeq
where $\beta_i(u) \equiv \kappa (\partial u_i /\partial \kappa)_{g^o}$.
One can then verify that the following equation holds:
\beql{rr59}
    \Gamma_R(\rho k_i, u, \kappa)
    =
    \rho^\alpha
    \Gamma_R(k_i,u(\rho), \kappa)\,,
\eeq
where
$
    \rho\, {du_i(\rho) /d \rho} = \beta_i(u)
$ and $\alpha$ is the bare dimension of the vertex function.
To obtain the IR behavior of a vertex function, it suffices
to scale it with its bare dimension $\alpha$ and
to substitute the new value of the running coupling $u(\rho)$.
In particular, when the normalization condition is
$
    \left.{\Gamma}_R(k, \kappa)\right|_{k=\kappa}
    = g_i = \kappa^{\alpha_i} u_i(1)
$
we obtain:
\beql{rr61}
    {\Gamma}_R(\rho k_i, u , \kappa)
    =
    \kappa^{\alpha_i} \rho^{\alpha_i} u_i(\rho)\,.
\eeq
In this way, if near the fixed point the running coupling is flowing to zero
with an exponent $y_i$
[$
    u_i(\rho)\sim \rho^{y_i}
$],
the vertex part will have the following IR behavior:
\beql{rr63}
    {\Gamma}_R(k) \sim k^{\alpha_i+y_i}\,.
\eeq
It is thus simple to convert the $\rhoo$-behavior of the running couplings
into the $k$-behavior of the relative vertex functions.
The simple relation between the renormalized couplings and the vertex
functions has been used to identify the $k\rightarrow 0$ limit of
$\Gamma_{ij}(k)$ given before \refe{smallrho}.


\subsection{Additional constraints on the running couplings}
\label{addconst}

We now proceed to write four RG equations for the left
four couplings: $v_{ll}$, $w_{lt}$, $u_{tt}$, and $z_{tt}$.
All other couplings can be readily obtained from them.
Indeed, the condensate $\psi_{lo}$ entering the WI
is the exact condensate density which is not scaling.
The other marginal couplings
$v_{ltt}$ and $v_{tttt}$ have been eliminated in favour
of $v_{ll}$ through Eqs. \refe{r87} and \refe{r88}.

It turns out that also the four left running couplings
left are not renormalizing independently.
As it was argued at the end of Section \ref{localWI}, we
expect that the combinations of couplings
$z_{tt}$, $w_{lt}/v_{ll}$, and
$c^2(\rhoo)= v_{ll} z_{tt}/(v_{ll} u_{tt}+w_{lt}^2)$
are invariant of the RG flow.
Here, we show how these properties for $w_{lt}/v_{ll}$ and $c$ can
be proved by analyzing the PT.
These identities are shown
to hold at one-loop level in Appendix \ref{AppRG1loop}.
The coupling $z_{tt}$ has to remain constant for the argument given
after \refE{r98}.
The absence of divergences in this case
follows directly from the WI and  power counting.
We now proceed to prove the other two identities by analysis of the
PT to all orders.

The coupling $w_{lt}$ can be identified with $v_{ll}$
via the WI (\ref{r79}) and (\ref{r80}),
which relate $w_{lt}$  to  $\Gamma_{tt;0}$ and
$v_{ll}$ to $\Gamma_{ltt}$, respectively.
By inspection of the leading singular terms to all
orders in perturbation theory,
$\Gamma_{tt;0}$ and $\Gamma_{ltt}$ are then found to be proportional
to each other.
As a matter of fact, the only way in which we can make an insertion of
the longitudinal ($l$) leg is to use the bare interaction term
$v_{ltt}$, since it connects the external longitudinal leg to two
internal transverse legs.
It is then clear that this corresponds to performing a density
insertion with the most relevant term (since
$|\psi|^2=\psi_t^2+\psi_l^2+2 \psi_{lo} \psi_t +\psi_{lo}^2$), apart
from the value of the bare external insertion that is different from
the bare $v_{ltt}$.
(See Appendix \ref{AppCoupling} for a discussion
on the density fields and the associated couplings.)
Diagrammatically this reads:
\beq
\Gamma_{tt;0} =
\!\!\!\parbox{13mm}{
\begin{fmfgraph*}(13,18)
  \fmftop{0} \fmfbottom{t1,t2}
  \fmfblob{.5w}{G}
  \fmf{gtt}{t1,G}  \fmf{gtt}{t2,G}
  \fmf{photon,label=$0$}{G,0}
\end{fmfgraph*}}
\sim
\parbox{13mm}{
\begin{fmfgraph*}(13,18)
  \fmftop{0} \fmfbottom{t1,t2}
  \fmfrpolyn{shaded,tension=.5}{G}{4}
  \fmf{gtt}{t1,G4}  \fmf{gtt}{t2,G3} \fmf{photon,label=$0$}{g,0}
  \fmf{gtt,left=.5,tension=.5}{G1,g}
  \fmf{gtt,right=.5,tension=.5}{G2,g}
  \fmfv{label=$\gamma_{tt;0}$,label.angle=0,label.dist=.2w}{g}
  \fmfdot{g}
\end{fmfgraph*}}
\sim
\parbox{13mm}{
\begin{fmfgraph*}(13,18)
  \fmftop{0} \fmfbottom{t1,t2}
  \fmfrpolyn{shaded,tension=.5}{G}{4}
  \fmf{gtt}{t1,G4}
  \fmf{gtt}{t2,G3} \fmf{gll,label=$l$}{g,0}
  \fmf{gtt,left=.5,tension=.5}{G1,g}
  \fmf{gtt,right=.5,tension=.5}{G2,g}
  \fmfv{label=$v_{ltt}$,label.angle=0,label.dist=.2w}{g}
  \fmfdot{g}
\end{fmfgraph*}}
\sim
\parbox{13mm}{
\begin{fmfgraph*}(13,18)
  \fmftop{0} \fmfbottom{t1,t2}
  \fmfblob{.5w}{G}
  \fmf{gtt}{t1,G}  \fmf{gtt}{t2,G}
  \fmf{gll,label=$l$}{G,0}
\end{fmfgraph*}}
\!\!\!=
\Gamma_{ttl}
\eeq
We can write the chain of identities in the following way
(by $\sim$ we mean the leading-order behavior as $k\rightarrow 0$):
\beqa
    \cho \Gamma_{ll}(k_0)
    &\stackrel{\rm (WI)}{\sim}&
       \Gamma_{ltt}(0,k_0)
    \stackrel{\rm (PT)}{\sim}
       {v_{ltt}\over \gamma_{tt;0}} \Gamma_{tt;0}(0,k_0)
       \nonumber \\
       & &\stackrel{\rm WI}{\sim}
       {v_{ltt} \over \gamma_{tt;0}}
       { \Gamma_{lt}(k_0) \over k_0}
    \label{t2}
\eeqa
where by $\gamma_{tt;0}$ we indicate the perturbative insertion
corresponding to $\Gamma_{tt;0}$ that in our case is $w_{lt}$
(see Appendix \ref{AppCoupling}).
In \refE{t2} we have set the spatial
part to zero exploiting the hypothesis of singular dependence
through only $k_0^2+\kv^2$.
We thus conclude that, at the leading order in the
divergent contributions:
\beq
    \Gamma_{ll}(k_0)
    \sim { v_{ll} \over w_{lt} } { \Gamma_{lt}(k_0) \over  k_0}
    \,.
    \label{t3}
\eeq
In this way, the logarithmic divergences entering in
the RG equation for $v_{ll}$ must also appear in the RG equation
for $w_{lt}$. If the exact (to all orders) RG equation for
$v_{ll}$ has the following form
\beq
    \rho {d v_{ll} \over d \rho} =  f(v_{ll},w_{lt},u_{tt})
    \,,
    \label{exp1}
\eeq
the resulting equation for $w_{lt}$ will be after a transient region:
\beq
    \rho {d w_{lt} \over d \rho} =
     f(\vd_{ll},w_{lt},u_{tt}) {w_{lt} \over v_{ll}}
    \, .
    \label{exp2}
\eeq
Dividing \refE{exp1} by \refE{exp2} we thus get:
\beq
    {d w_{lt} \over d v_{ll} }  =
    {w_{lt} \over \vd_{ll}}
    \label{exp3}
\eeq
impling
\beq
    {v_{ll}(\rho)\over w_{lt}(\rho)} = {\rm constant}
    \equiv C_1
    \,.
    \label{exp4}
\eeq

In a similar way,  the invariance
of $c(\rhoo)$  follows from
the exact connection between the
singular parts of
$\Gamma_{;00}$
and
$\Gamma_{ll}$,
associated  respectively with $u_{tt}$ and $v_{ll}$:
\beq
\Gamma_{;00} =
\!\!\!\parbox{13mm}{
\begin{fmfgraph*}(13,18)
  \fmftop{l1} \fmfbottom{l2}
  \fmfblob{.5w}{G}
  \fmf{photon,label=$0$}{l1,G}  \fmf{photon,label=$0$}{l2,G}
\end{fmfgraph*}}
\sim
\parbox{13mm}{
\begin{fmfgraph*}(13,25)
  \fmftop{l1} \fmfbottom{l2}
  \fmfrpolyn{shaded,tension=.1}{G}{4}
  \fmf{photon,label=$0$}{g1,l1}
  \fmf{photon,label=$0$}{g2,l2}
  \fmf{gtt,left=.5,tension=.5}{G3,g1}
  \fmf{gtt,right=.5,tension=.5}{G4,g1}
  \fmf{gtt,left=.5,tension=.5}{g2,G2}
  \fmf{gtt,right=.5,tension=.5}{g2,G1}
  \fmfv{label=$\gamma_{tt;0}$,label.angle=0,label.dist=.25w}{g1}
  \fmfv{label=$\gamma_{tt;0}$,label.angle=0,label.dist=.25w}{g2}
  \fmfdot{g1,g2}
\end{fmfgraph*}}
\sim
\parbox{13mm}{
\begin{fmfgraph*}(13,25)
  \fmftop{l1} \fmfbottom{l2}
  \fmfrpolyn{shaded,tension=.1}{G}{4}
  \fmf{gll,label=$l$}{g1,l1}
  \fmf{gll,label=$l$}{g2,l2}
  \fmf{gtt,left=.5,tension=.5}{G3,g1}
  \fmf{gtt,right=.5,tension=.5}{G4,g1}
  \fmf{gtt,left=.5,tension=.5}{g2,G2}
  \fmf{gtt,right=.5,tension=.5}{g2,G1}
  \fmfv{label=$v_{ltt}$,label.angle=0,label.dist=.25w}{g1}
  \fmfv{label=$v_{ltt}$,label.angle=0,label.dist=.25w}{g2}
  \fmfdot{g1,g2}
\end{fmfgraph*}}
\sim
\parbox{13mm}{
\begin{fmfgraph*}(13,18)
  \fmftop{l1} \fmfbottom{l2}
  \fmfblob{.5w}{G}
  \fmf{gll,label=$l$}{l1,G}  \fmf{gll,label=$l$}{l2,G}
\end{fmfgraph*}}
\!\!\!=
\Gamma_{ll}
\eeq
This procedure leads to the following asymptotic equation for $\ud_{tt}$
\beq
    \rho {d u_{tt} \over d \rho} =
    - f(v_{ll},w_{lt},u_{tt}) {w_{lt}^2 \over v_{ll}^2}
    \,.
    \label{exp5}
\eeq
Using Eq.~\refe{exp4} we obtain
$
du_{tt}/dv_{ll} = - C_1^{-2}
$
and
\beq
    \ud_{tt}(\rho) = -C_1^{-2}\,\vd_{ll}(\rho)+C_2 \,,
    \label{exp6}
\eeq
so that $u_{tt}+w^2_{lt}/v_{ll} = C_2$ and $c(\rho)$ is invariant
(cf. Eq.~\refe{s6}).
The values of the two constants $C_1$ and $C_2$ can be extracted
from the $\rho\rightarrow 0$ limit of the running couplings,
specifically from Eqs.\ \refe{s7} and \refe{s8}.
The flow of the four running couplings will satisfy thus these three
identities. We are left eventually with only one
independent running coupling,
for instance, $\vd_{ll}$.

The constraints found above for $v_{ll}$, $w_{lt}$, $z_{tt}$,
and $u_{tt}$ fix completely the IR behavior of two out of three
Green's functions, say  $\Gc_{lt}(k)$ and $\Gc_{tt}(k)$.
From \refE{r57} one can in fact say that the expressions
for $\Gc_{lt}(k)$ and $\Gc_{tt}(k)$ at vanishing  $v_{lt}$ and
$v_{tt}$ involve only the above combinations of running couplings,
which
do not scale and are related to physical quantities.
The resulting {\em exact} IR behavior is thus given by:
\beqa
    \Gc_{tt}(\kv,\omega) &=&
    -{c^2 m  n_o \over n} {1\over \omega^2+c^2 \kv^2}
    \label{t14}
    \\
    \Gc_{lt}(\kv,\omega) &=&
    -{m c^2 \over 2 n} {dn_o \over d\mu}
    {\omega \over \omega^2+c^2 \kv^2}
    \,.
    \label{t15}
\eeqa
This result is valid at all orders of perturbation theory and does not
depend on the scaling of $v_{ll}$.
The Goldstone mode is accordingly stabilized only by symmetry
requirements, irrespective of the behavior of $v_{ll}$.
The vanishing of $v_{ll}$ ($\sim \Sigma_{12}$) is no longer changing the
IR behavior of $\Gc_{tt}$ and $\Gc_{lt}$.
As discussed in the Introduction, the IR behavior of $\Gc_{ll}$ is
expected to depend on dimensionality, and thus cannot be fixed by
symmetry requirements only.


\subsection{One-loop equation for $v_{ll}$}
\label{oneloop}


To proceed further we need to perform a specific calculation
of the RG equation for one coupling,
for instance $v_{ll}$.
At the one-loop level, only one diagram contributes
to $\Gamma_{ll}$ (see Appendix \ref{App1loop}).
(Derivation of the other one-loop equations is not required. For
comparison, we give the full set of one-loop equations in Appendix
\ref{AppRG1loop} for the marginal running couplings entering the
propagators.)
With dimensional regularization and after the $\epsilon$ expansion
has been carried out, $\Gamma_{ll}$ reads:
\beq
    \Gamma_{ll}(k) = v_{ll} -
    {v_{ll}^2 \kappa^{-\epsilon}
    \over
    2 \psi_{lo}^2 z_{tt}^2 \epsilon}
    \label{t6}
\eeq
where $\kappa^2=\kv^2+k_o^2$ and the factor $K_d=2/(4\pi)^{d/2}\Gamma(d/2)$
has been absorbed in the definition of the running coupling.
We made use of Eq.~\refe{r87} to connect $v_{ltt}$ to $v_{ll}$.
To set up the RG equations, we need to define dimensionless running
couplings (see Section \ref{RGApproach}).
Since all  quadratic running couplings are already dimensionless
(at any dimension), only the condensate density $\psi_{lo}$ needs to be
scaled by a factor $\kappa^{\epsilon/2}$:
\beq
    \bar\psi_{lo} = \psi_{lo} \kappa^{\epsilon/2}
    \,.
    \label{t7}
\eeq
Note that the powers of $c_o$ have already been eliminated
in $\psi_{lo}$ which is redefined like $c_o^{-3/2}\psi_{lo}$.
By minimal subtraction of the $1/\epsilon$ pole one obtains the
following RG equation for $v_{ll}(\rhoo)$:
\beq
    \rhoo {d v_{ll}\over d \rhoo}
    =
    {v_{ll}^2 \over 2 \psi_{lo}^2(\rho) z_{tt}^2}
    \, ,
    \label{t8}
\eeq
where $\psi_{lo}(\rho)=\psi_{lo}(1)\rhoo^{\epsilon/2}$ exactly
while $z_{tt}$ is invariant.
We stress that the scaling of $\psi_{lo}$ is a trivial effect of
its dimension and the condensate density is not renormalizing.
An alternative procedure would be to fix the condensate density
and use scale dependent WI in order to connect the running
couplings $v_{ll}$, $v_{ltt}$, and $v_{tttt}$.
The solution of \refE{t8} is straightforward and depends on dimensionality.
For $d=3$, $\psi_{lo}$ is constant and
\beq
    {v_{ll}(\rhoo)\over v_{ll}(1) }
    =
    \left[
    1 - \displaystyle
    {v_{ll}(1) \over 2 \psi_{lo}(1)^2 z_{tt}(1)^2} \ln \rhoo
    \right]^{-1}
    \,.
    \label{t9}
\eeq
For $d<3$ we obtain instead:
\beq
    { v_{ll}(\rho) \over v_{ll}(1)}=
    \left[
    1 +
    {v_{ll}(1) \over 2 \psi_{lo}(1)^2 z_{tt}(1)^2}
    {\rhoo^{-\epsilon}-1 \over \epsilon}
    \right]^{-1}
     \, .
    \label{t10}
\eeq
We recall that the RG equations are valid only after a transient
region, so that the boundary conditions for \refE{t8} should
involve the running couplings at some intermediate value $0<\bar\rho<1$.
For simplicity, we write the solution of this equation in terms
of the coupling at $\rho=1$. Since the results do not depend on
the boundary conditions (provided the sign is not changed), we
assume that the $\rho=1$ is the point where the RG equations start
being accurate.

In both cases \refe{t9} and \refe{t10}
the coupling $v_{ll}$ is flowing to zero, implying
the vanishing of $\Sigma_{12}(k\rightarrow 0)$
(here $\Sigma_{12}=[\Gamma_{ll}-\Gamma_{tt}+i(\Gamma_{lt}+\Gamma_{tl})]/4$).
At the same time, the constraint on the other couplings implies
that $w_{lt}(\rho)$ vanishes like $v_{ll}(\rho)$, which from
\refE{s6} gives that
$u_{tt}(\rho)$ flows to $z_{tt}(\rho) c^2(\rho)/c_o^2$.
Since both $c(\rho)$ and $z_{tt}(\rho)$ are constants of motion of the
RG flow, $u_{tt}(\rho) \rightarrow z_{tt}(1)$.
We can thus find the asymptotic expression for the longitudinal Green's
functions as follows:
\beq
    \Gc_{ll}(k)
    =  -{1\over v_{ll} u_{tt}}
      {u_{tt}\, k_0^2 + z_{tt} \kv^2
      \over
      k_0^2 + (c^2/c_o^2) \kv^2}
      \rightarrow
      -{1\over v_{ll}}
      \,.
    \label{t11}
\eeq
According to the scaling \refe{rr63} of the RG,  this implies
that
\beq
    \Gc_{ll}(\rho k) \sim {1\over v_{ll}(\rho)}
    \,.
    \label{t12}
\eeq

The above result, together with Eqs.\ \refe{t14} and \refe{t15}, gives
the complete IR behavior for all one-particle Green's functions.
By using the results of Section \ref{SecIII} on the finite value of
the small $k$-limit of the vertex functions, it is possible to rewrite
the result \refe{t12} in terms of thermodynamical quantities:
\beqa
    \Gc_{ll}(\kv,\omega) &=& {c^2 \over 8 n_0 n}
    \left({dn_0\over d\mu}\right)^2
    {\omega^2 \over \omega^2+ c^2 \kv^2}
    \nonumber \\
    && \quad
    +\left\{
    \begin{array}{lc}
    { \disp c_o n_o \over \disp 64 \pi^2 n^2} \ln(k/\Lambda) & (d=3)\\
    \disp -{c_o n_o K_{4-\epsilon} \over 8 n^2 \epsilon k^\epsilon}
    &
    (1<d<3)
    \end{array}
    \right.
    \label{t13}
\eeqa
where $\Lambda$ is an ultraviolet cutoff of the order of the sound velocity
(corresponding to the region where the RG starts to be accurate)
and  $c_o$ is the sound velocity at the scale $\Lambda$.
After that point, $c(\rho)$ is, to a good accouracy, an invariant of the
RG flow, so that we can substitute the fixed point value $c_o=c$.

We note that for $d=1$ ($\epsilon=2$) the transverse and longitudinal
divergences become identical while
the condensate vanishes.
It is thus expected that transverse and longitudinal fluctuations
become symmetric since it is no longer possible even to define
longitudinal and  transverse directions.
The main assumptions of our approach will thus break down,
indicating the existence of a lower critical dimension.
Nevertheless, since the results are valid exactly for $1<d<3$, as
said already, the predicted IR behavior approaches the correct result
for $d \rightarrow 1$.\cite{Popov}

The usual Green's functions can be readily obtained from
Eqs.\ \refe{t14}, \refe{t15} and \refe{t13}:
\begin{mathletters}
\beqa
    \Gc_{11}(\kv,\omega) &=&
    -{c^2 m  n_o \over n} {1\over \omega^2+c^2 \kv^2}
    + {m c^2 \over n} {dn_o \over d\mu}
    {i\omega \over \omega^2+c^2 \kv^2}
    \nonumber \\
    &&\quad + \Gc_{ll}(\kv,\omega)
    \,,
    \label{t16}
    \\
    \Gc_{12}(\kv,\omega) &=&
    {c^2 m  n_o \over n} {1\over \omega^2+c^2 \kv^2}
    + \Gc_{ll}(\kv,\omega)
    \,,
    \label{t17}
\eeqa
\label{t16-17}
\end{mathletters}
where we have restored the canonical dimensions and reintroduced the
mass $m$.
The leading order of the expressions \refe{t16-17} (i.e., the
first term) coincides with that found by Gavoret and
Nozi\`eres.\cite{GN}
The additional terms proportional to $\Gc_{ll}$ give the next-to-leading
contribution to the Green functions coming from the divergent
diagrams.  These diverging sub-leading contributions were identified
for the first time in Ref. \onlinecite{Nepo} without calculating the
coefficients.
Later, by improving the Bogoliubov approximation and exploiting the
analogy with spin-wave theory, Weichman found an explicit expression
for $\Gc_{ll}$ valid for the weak-coupling Bose gas.\cite{Weich88}
Assuming a Landau quantum-hydrodynamic Hamiltonian, the authors of
Ref.~\onlinecite{Stringari} have derived the same expression given by
\refe{t13} for the longitudinal correlation function.


\subsection{Extension to all orders of PT}


Although the IR behavior of $\Gc_{ij}$ has been found only at the
one-loop order, we will show now that it is actually correct to
all orders.
We begin by studying the nature of the fixed point.
The scaling of the running couplings near the fixed
point is given by Eqs.~\refe{t9} and \refe{t8} for $v_{ll}$
and by the WI for $v_{ltt}$ and $v_{tttt}$:
\beqa
    v_{ltt}(\rhoo)
    &=&
    v_{ll}(\rhoo)/\psi_{lo}(\rhoo)
    \sim
    v_{ll}(\rho)\, \rho^{-\epsilon/2}
\\
    v_{tttt}(\rho)
    &=&
    3 \, v_{ltt}(\rho)/\psi_{lo}(\rho)
    \sim v_{ll}(\rho)\, \rho^{-\epsilon}
    \label{rr76}\,.
\eeqa
One thus finds that $v_{ltt}$ always vanishes near the fixed point.
For $d=3$  $v_{tttt}$ flows also to zero, while for $d<3$ it
flows to a finite fixed point value $v_{tttt}^*$.
We can thus draw one important conclusion, namely
that for $\epsilon=0$ the theory is
asymptotically free.
Marginal interactions become marginally irrelevant (they vanish like
$1/\ln\rho $) and the results obtained at one loop are exact, in
agreement with Benfatto calculations.\cite{Benfatto}

Unfortunately, for $\epsilon>0$ the theory is not free
since $v_{tttt}$ flows towards a finite value.
This fact may, in principle, invalidate the results found at
one-loop level, because higher-order terms are not
negligible near the fixed point.
Note, however, that contrary to what happens in critical phenomena,
the asymptotic behavior of $v_{ll}$ does not depend on the position
of the fixed point ($v_{tttt}^*$) but only on its very existence.
This can be verified from \refE{rr76}.
When $v_{tttt}(\rhoo\rightarrow 0)$ is finite,
$v_{ll}(\rhoo) \sim \rhoo^{\epsilon}$ and the exponent does not
depend on the values of interactions.
For this reason it is possible to establish the scaling of $v_{ll}$
without a {\em precise} knowledge of the corresponding RG equation.
Actually its ``form'' is enough, as we shall argue and explain
in the following by analyzing the PT.
One has to prove either that $v_{ll}(\rho) \sim \rho^\epsilon$,
or the existence of the fixed point, since the two statements
are equivalent.
We will prove the former statement in the following.

Let us consider the following two-loops diagrams contributing
to $\Gamma_{ll}$:
\beq
\parbox{20mm}{
\begin{fmfgraph*}(20,10)
  \fmfleft{l} \fmfright{r}
  \fmftop{t} \fmfbottom{b}
  \fmf{phantom,tension=5}{t,ti}  \fmf{phantom,tension=5}{b,bi}
  \fmf{gtt,left=.3,tension=.5}{v1,ti,v2}
  \fmf{glt,left=.3,tension=.5}{bi,v1}
  \fmf{gtt,right=.3,tension=.5}{bi,v2}
  \fmf{glt,tension=.2}{ti,bi}
  \fmf{gll,tension=1.}{l,v1}
  \fmf{gll,tension=1.}{r,v2}
  \fmfdot{v1,v2,bi,ti}
\end{fmfgraph*}}
\,,\quad
\parbox{20mm}{
\begin{fmfgraph*}(20,10)
  \fmfleft{l} \fmfright{r}
  \fmftop{t} \fmfbottom{b}
  \fmf{phantom,tension=5}{t,ti}  \fmf{phantom,tension=5}{b,bi}
  \fmf{gtt,left=.3,tension=.5}{v1,ti,v2}
  \fmf{gtt,right=.3,tension=.5}{v1,bi,v2}
  \fmf{gll,tension=.2}{ti,bi}
  \fmf{gll,tension=1.}{l,v1}
  \fmf{gll,tension=1.}{r,v2}
  \fmfdot{v1,v2,bi,ti}
\end{fmfgraph*}}
\, , \,
\parbox{20mm}{
\begin{fmfgraph*}(20,10)
  \fmfleft{l} \fmfright{r}
  \fmf{gtt,left=.8,tension=.9}{v1,vc} \fmf{gtt,left=.8,tension=.9}{vc,v2}
  \fmf{gtt,right=.8,tension=.9}{v1,vc}  \fmf{gtt,right=.8,tension=.9}{vc,v2}
  \fmf{gll,tension=4.}{l,v1}
  \fmf{gll,tension=4.}{r,v2}
  \fmfdot{v1,v2,vc}
\end{fmfgraph*}}
\,.
\label{diadue}
\eeq
These diagrams, once expanded in $\epsilon$, develop
$1/\epsilon$ poles, and thus they contribute to the equation
for $v_{ll}(\rho)$.
Without explicit calculations, we can infer the form of this
contributions.
We have seen that for $\rho \rightarrow 0$, $\Gc_{tt}(k)$ and
$\Gc_{lt}(k)$ do not depend on the explicit scaling of $v_{ll}$, thus
they will not contribute to \refe{diadue} with factors
$v_{ll}$ (at least at leading order as $\rho\rightarrow 0$).
This means, in particular, that the first diagram
contributes to the beta-function only with a term
proportional to $v_{ltt}^4$, coming from the vertices.
Its contribution is thus negligible with respect to the one-loop
contribution \refe{t8}.
These classes of diagrams do not affect the RG equations.

Let us consider now the second diagram in \refe{diadue},
for which  the situation is different since $\Gc_{ll}$ is proportional to
$1/v_{ll}(\rho)$, but each internal $\Gc_{ll}$ must appear in
combination with two $v_{ltt}$.
From the scaling form at one-loop level, that we assume valid
for the moment, we find that as $\rho\rightarrow 0$:
\beq
    v_{ltt}^2(\rho) \Gc_{ll}(k)
    \sim
    -{v_{ltt}^2(\rhoo) \over \vd_{ll}(\rhoo)}
    \sim
    -{v_{tttt}^* \over 3}
    \label{rr83}
    \,.
\eeq
Taking into account \refE{rr83} this implies that the form the
contribution of the second diagram in \refe{diadue} is of the same form
of \refe{t8}, {\em i.e.}, it has exactly the same $\rho$ dependence.
It does thus not change the form of the solution for $v_{ll}$,
even though it can give a multiplicative factor in the final scaling.

The last diagram in \refe{diadue} has a behavior similar to
the second one.
It depends on $v_{ll}$ only through
the two $v_{ltt}$ connected to the external $\Gc_{ll}$.
The internal part gives a finite contribution in $\rho$  and again,
once added to \refe{t8}, it does not change the scaling of $v_{ll}$.
We thus conclude that the contributions of {\em all} two-loops diagrams
to \refe{t8} do not change the IR behavior of $v_{ll}$.

The same arguments can be extended at higher order in loops.
The structure of the PT remains the same.
The presence of even only one $\Gc_{lt}$ is enough to suppress
the diagram due to the $v_{ltt}$ non compensated by any $1/v_{ll}$.
The $\Gc_{ll}$ appears always in combination with two $v_{ltt}$,
giving a finite non-dangerous term.
All diagrams thus are either negligible, or give
a finite contribution for $\rho\rightarrow 0$
multiplied by two $v_{ltt}$ coming from the
two vertices connected to the external lines.

The RG equation for $v_{ll}$ to all orders has thus the following
simple form:
\beqa
    \rho {d v_{ll} \over d \rho}
    &=&
    {v_{ll}^2 \over \psi_{lo}^2(\rho)}
    \left[
    A_0
    +A_1 v_{tttt}^*
    +A_2 (v_{tttt}^*)^2
    + \dots
    \right]
    \,.
    \label{allorders}
\eeqa
Provided the sum of the series does not change sign with
respect to the sign of the first term, for $\rho\rightarrow 0$
we obtain:
\beq
    v_{ll}(\rho)
    =
    {
    \psi_{lo}^2(a)
    \over A(v_{tttt}^*)
    }
    \epsilon \rho^{\epsilon}
    \label{rr87}
\eeq
where $a<1$ is the value of $\rho$ for which the leading behavior
established by \refe{allorders} holds
and $A(v_{tttt}^*)$ is the sum of the series of \refE{allorders}.
This concludes our proof.
Higher-order diagrams do not change the IR behavior,
as they only can affect the finite coefficients.

We have thus proved that the scaling \refe{t10} holds apparently
for any value of $\epsilon>0$.
One has to recall that the procedure breaks down
at $d=1$, since the condensate vanishes in that case,
corresponding to  $\epsilon<2$.
In conclusion, we have proved that the IR behavior found at one loop
is maintained for $1 < d \leq 3$.

\subsection{A non-trivial fixed point without anomalous dimensions}

We have shown that the fixed point is characterized by a finite value
of the interacting constant among transverse fluctuations, similarly to
what happens for the $\phi^4$ theory.
We have also shown that no anomalous dimension appears in the
transverse correlation function.
This is clearly due to the constraints originating from the
WI.
The question is now how these constraints can be
implemented in the structure of the perturbation theory.
Actually, the same diagrams responsible for the anomalous
dimension in the $\phi^4$ theory are also present here
in $\Gamma_{tt}$.
For instance:
\beq
\parbox{25mm}{
\begin{fmfgraph*}(25,10)
  \fmfleft{l} \fmfright{r}
  \fmf{gtt,left=.5, tension=.3}{v1,v2}
  \fmf{gtt,right=.5,tension=.3}{v1,v2}
  \fmf{gtt,tension=.3}{v1,v2}
  \fmf{gtt,tension=3.}{l,v1}
  \fmf{gtt,tension=3.}{r,v2}
  \fmfdot{v1,v2}
\end{fmfgraph*}}
\quad .
\eeq
These diagrams, however, do not contribute in practice because
other diagrams contribute
with opposite sign and remove the divergence.
This can be realized to all orders by looking at the form of
\refE{rr83}.
The structure $\Gc_{ll} v_{ltt}^2$ acts as an effective
$v_{tttt}$ interaction mediated by the longitudinal field.
According to \refE{rr83}, it has the opposite sign with respect to the
fixed point interaction $v_{tttt}^*$.
All contributions to the effective interaction are given
by the following graphical equation:
\beq
    v^{eff}_{tttt}=
    \parbox{12mm}{
    \begin{fmfgraph*}(10,8)
    \fmfleft{l1,l2}\fmfright{r1,r2}
    \fmf{gtt}{r1,v,r2}\fmf{gtt}{l1,v,l2}
    \fmfdot{v}
    \end{fmfgraph*}}
    +
    \parbox{12mm}{
    \begin{fmfgraph*}(12,10)
    \fmfleft{l1,l2}\fmfright{r1,r2}
    \fmf{gtt}{r1,v1,r2} \fmf{gtt}{l1,v2,l2}
    \fmf{gll}{v1,v2}
    \fmfdot{v1}\fmfdot{v2}
    \end{fmfgraph*}}
    +
    \parbox{12mm}{
    \begin{fmfgraph*}(12,10)
    \fmfleft{l1,l2}\fmfright{r1,r2}
    \fmf{gtt}{r1,v1,l1} \fmf{gtt}{r2,v2,l2}
    \fmf{gll}{v1,v2}
    \fmfdot{v1}\fmfdot{v2}
    \end{fmfgraph*}}
    +
    \parbox{12mm}{
    \begin{fmfgraph*}(12,10)
    \fmfleft{l1,l2}\fmfright{r1,r2}
    \fmf{gtt,tension=.2}{r1,v1} \fmf{gtt}{v1,l2}
    \fmf{gtt}{r2,v2} \fmf{gtt,tension=.2}{v2,l1}
    \fmf{gll,tension=.1}{v1,v2}
    \fmfdot{v1}\fmfdot{v2}
    \end{fmfgraph*}}
    \,.
    \label{veff}
\eeq
Evaluating the diagrams entering \refE{veff}, one finds that at the
fixed point $v_{tttt}^{eff}$ vanishes due to the exact cancellation
of the fixed point $v_{tttt}^*$ with the interaction mediated by
$\Gc_{ll}$.

This explains why no $\phi^4$-like divergences appear in the
contribution to $\Gamma_{tt}$. This does not mean, however, that the theory
is free at the fixed point.
As a matter of fact, $v^{eff}_{tttt}$ is not defined in terms of
one-particle irreducible diagrams for $\Gamma_{tttt}$.
Thus the above cancellation is only partial in some diagrams for
$\Gamma_{ll}$.
This can be seen by expanding in a  typical diagram for $\Gamma_{ll}$
the definition of $v^{eff}_{tttt}$
(represented here by the dashed blob):
\beqa
\parbox{20mm}{
\begin{fmfgraph*}(20,10)
  \fmfleft{l} \fmfright{r}
  \fmf{gtt,left=.8,tension=.9}{v1,vc} \fmf{gtt,left=.8,tension=.9}{vc,v2}
  \fmf{gtt,right=.8,tension=.9}{v1,vc}  \fmf{gtt,right=.8,tension=.9}{vc,v2}
  \fmf{gll,tension=4.}{l,v1}
  \fmf{gll,tension=4.}{r,v2}
  \fmfdot{v1,v2,vc} \fmfblob{.1w}{vc}
\end{fmfgraph*}}
&=&
\parbox{20mm}{
\begin{fmfgraph*}(20,10)
  \fmfleft{l} \fmfright{r}
  \fmf{gtt,left=.8,tension=.9}{v1,vc} \fmf{gtt,left=.8,tension=.9}{vc,v2}
  \fmf{gtt,right=.8,tension=.9}{v1,vc}  \fmf{gtt,right=.8,tension=.9}{vc,v2}
  \fmf{gll,tension=4.}{l,v1}
  \fmf{gll,tension=4.}{r,v2}
  \fmfdot{v1,v2,vc}
\end{fmfgraph*}}
\,+2 \,
\parbox{20mm}{
\begin{fmfgraph*}(20,10)
  \fmfleft{l} \fmfright{r}
  \fmftop{t} \fmfbottom{b}
  \fmf{phantom,tension=5}{t,ti}  \fmf{phantom,tension=5}{b,bi}
  \fmf{gtt,left=.3,tension=.5}{v1,ti,v2}
  \fmf{gtt,right=.3,tension=.5}{v1,bi,v2}
  \fmf{gll,tension=.2}{ti,bi}
  \fmf{gll,tension=2.}{l,v1}
  \fmf{gll,tension=2.}{r,v2}
  \fmfdot{v1,v2,bi,ti}
\end{fmfgraph*}}
\nonumber\\
&& + \,
\parbox{20mm}{
\begin{fmfgraph*}(20,10)
  \fmfleft{l} \fmfright{r}
  \fmf{gtt,left=.8,tension=.2}{v1,vc1}
  \fmf{gll,tension=.4}{vc1,vc2}
  \fmf{gtt,left=.8,tension=.2}{vc2,v2}
  \fmf{gtt,right=.8,tension=.2}{v1,vc1}
  \fmf{gtt,right=.8,tension=.2}{vc2,v2}
  \fmf{gll,tension=2.}{l,v1}
  \fmf{gll,tension=2.}{r,v2}
  \fmfdot{v1,v2,vc1,vc2}
  \,.
\end{fmfgraph*}}
\eeqa
The above diagram vanishes, but the last diagram does not contribute to
$\Gamma_{ll}$ since it is not one-particle irreducible.
Thus, $v_{tttt}$ does affect the equation for $v_{ll}$ even if it gives
only finite renormalization, as shown before.
On the other side, one can verify that all $v_{tttt}$ terms appearing
at all orders in $\Gamma_{tt}$ are exactly canceled by the
mediated interaction.
We have thus explicitly shown how the diagrams responsible
for the appearance of the anomalous dimension in $\phi^4$ theory
get eliminated in the present case.
This is the counterpart of what happens in the phase-amplitude variables,
where the integration of the amplitude
fields reconstructs an effective full-gauge-invariant action.


\subsection{IR behavior of the the response functions}
\label{result}

In Section \ref{oneloop} we have found the IR behavior of
the quadratic Green's functions.
Only $\Gc_{ll}$ is affected by the nontrivial scaling
of the four quadratic running couplings.
We found that the form of the spectrum is unaffected by
the RG flow.
In the same way, we find that the physical response functions,
specifically the density-density, density-current, and current-current
response functions ($\Gc_{\mu,\nu}(k)$) depend only
on RG-invariant combinations of the running couplings.
This is not surprising, since the expressions found
for these functions in the limit $k\rightarrow 0$
by Gavoret and Nozi\`eres depend only on two
finite physical quantities, $n$ and $c^2$.
As  an example, we consider explicitly the density-density
correlation function $\Gc_{;00}$.
We obtain its expression from the general relation:
\beq
    \Gc_{;\nu\mu}(p)
    =
    \Gamma_{;\nu \mu}(p)
    +\Gamma_{j;\nu}(-p)\Gc_{ij}(-p)\Gamma_{i;\mu}(p)
    \label{e5.45}
\eeq
with $\mu=\nu=0$. The scaling of the composite $\Gamma$ functions
has been linked to the quadratic running couplings through
the WI \refe{r94} and \refe{r96}:
\[
    \Gamma_{l;0}(\kv,\omega)
    \sim
     w_{lt} n_0^{1/2}
    \quad,\qquad
    \Gamma_{t;0}(\kv,\omega)
    \sim
     u_{tt} n_0^{1/2} \omega
    \,.
\]
Substituting this expression into \refE{e5.45} and using
$u_{tt} \rightarrow -\Gamma_{;00}/n_o$ in the IR, we recover
the Gavoret-Nozi\`eres result (with the mass reinserted explicitly):
\beq
    \Gc_{;00}(p)
    =
    -{ n \kv^2/(n_0m)  \over \omega^2+c^2 \kv^2} \,.
\eeq
No anomalous terms appear. The same is true for the
other response functions.

\section{Conclusions}
\label{SecVI}

Infrared divergences plague perturbation theory of the interacting
Bose gas without the presence of a true physical instability for the
system.
This means that IR divergences must disappear from the final
physical quantities, like the thermodynamical derivatives.
However, it is
not trivial to control the divergent parts of the perturbation
theory whithin approximated shemes.
Divergences are due to the strong fluctuations of the
phase of the broken-gauge-symmetry order parameter.
In this paper, we have set up a RG approach to treat
these singularities in a systematic way.
Since they must disappear in the final results, most divergences are
actually related.
We implemented these connections through Ward
identities.
We found that WI are sufficient to guarantee the stability of
the superfluid phase with respect to phase fluctuations.
As a consequence, the sound velocity of the macroscopic and microscopic
excitations is an invariant of the RG flow.

Nevertheless the divergences are not totally ineffective.
The WI allow to reduce the set of independent running couplings
to a single one, for which we have derived a flow equation at the
one-loop order.
Within this approximation, we found  that the
theory becomes free at $d=3$, {\em i.e.}, all running couplings flow
to zero in agreement with Benfatto result.\cite{Benfatto}
One-loop order proves thus enough to obtain the exact result at $d=3$, that
gives a logarithmically divergent longitudinal correlation function
$\Gc_{ll}$.
For $1<d<3$ the fixed point is characterized by a finite
value of the coupling $v_{tttt}$  among transverse fluctuations,
being a truly interacting fixed point.
Even in this case we found that the one-loop result gives the
correct exponent for $\Gc_{ll}$.
Higher-order diagrams are not
negligible like as in the $d=3$ case, but they contribute
like the one-loop ones, and thus they do not change the IR behavior.
Concerning the two other correlation functions $\Gc_{lt}$ and $\Gc_{tt}$,
we found that their IR behavior is determined by WI and no
anomalous dimension appears.
This result can be surprising for the transverse one, since at the
fixed point the interaction is finite and an anomalous dimension
could appear.
We showed, however, how the structure of the PT adapts in a nontrivial
way near the fixed point, in order to keep
$v_{tttt}$ finite and fulfill WI.
The transverse field is effectively free once one adds to
$v_{tttt}$ the interaction mediated by the longitudinal fields.

In conclusion, our approach has enabled us to prove, within a coherent frame,
the disappearance of the divergences in the physical quantities
while keeping a firm control on singularities during the procedure.
The nontrivial evolution of the structure of the propagator
from large momenta (Bogoliubov) to small momenta (fixed point)
can be  readily followed in terms of the flow of the running couplings.
This approach can be useful, in general, to study stable systems where
singularities appears due to a broken continuous
symmetry without true instabilities.

We acknowledge discussions with Ph. Nozi\'eres, G. Benfatto, and
M. Grilli.
F.P. gratefully acknowledges Europa Metalli S.p.A. for partial financial
support and the Statistical Mechanics and Complexity (SMC) center
in Rome for hospitality in the last stage of this work.

\appendix

\section{Gauge invariance constraints for the running couplings}

\label{AppCoupling}

The introduction in \refE{r53} of the couplings associated
with the monomials of the $\chi_{l/t}$ fields
may lead to some difficulties in defining gauge invariance itself.
Since the theory we are constructing must satisfy the
original gauge invariance defined in \refE{r60} in terms of the
$\psi$ fields, we need to specify the the equivalent
transformation for the action \refe{r53} in terms of the new fields.
In this way, we can establish in a different way a connection
among the bare couplings of the theory.

To discuss this point we proceed by steps.
We begin with the global part of the
action and introduce the following simplified notation for the couplings:
\beq
    S =
    \sum_{n,m} {v_{n,m} \over m! n!} \psi_l^n(x) \psi_t^m(x)
    + \lambda_l (\psi_l+\alpha_l) + \alpha_t (\psi_t+\alpha_t)
    \label{AA1}.
\eeq
In this expression, $\alpha_{l/t}$ are two {\em arbitrary} real numbers.
The global gauge invariance in terms of these fields reads:
\beq
    \left[{ \psi'_l(x) \atop \psi'_t(x)} \right]
    =
    {\bf R}(\chi)
    \left[{ \psi_l(x)+\alpha_l \atop \psi_t(x)+\alpha_t} \right]
    -\left[{ \alpha_l \atop \alpha_t} \right]
    \label{AA2}
\eeq
where ${\bf R}$ is defined in \refE{r62} and the external
fields change as follows:
\beq
    \left[{ \lambda'_l(x) \atop \lambda'_t(x) } \right]
    =
    {\bf R} \left[{ \lambda_l(x) \atop \lambda_t(x) } \right]
    \label{AA3}
    \,.
\eeq
Note that the transformation \refe{AA2} depends on $\alpha_{l/t}$.
This is due to the fact that
the new action is no longer invariant under
rotations around the origin in the space $(\psi_l,\psi_t)$,
but rather around the point $(-\alpha_l,-\alpha_t)$.
It is also clear the the original action is manifestly
gauge invariant, since it is a sum of gauge invariant terms
(essentially $|\psi|^{2n}$).
By contrast, the action \refe{AA1} is invariant under the transformations
\refe{AA2}-\refe{AA3} only for specific values of the set of
couplings $\{ v_{nm}\}$.
It is easy to find these constraints  by implementing the transformation
\refe{AA2}-\refe{AA3} infinitesimally:
\beq
    \left\{
    \begin{array}{rcl}
    \delta\psi'_l(x) &=& -(\psi_t(x)+\alpha_t) \delta \chi
    \\
    \delta\psi'_t(x) &=& -(\psi_l(x)+\alpha_l) \delta \chi
    \end{array}
    \label{AA4}
    \right. .
\eeq
Imposing the invariance of $S$ given by \refe{AA1}
under the transformation \refe{AA4} gives
the following set of equations (with $\alpha_t=0$):
\beq
    -m\, v_{n+1,m-1} +n\, v_{n-1,m+1} + \alpha_l \, v_{n,m+1} =0
    \,.
    \label{AA5}
\eeq
One can readily verify that these conditions coincide with
the WI for $k_i=0$,
the two procedures being equivalent.

At this point it is easy to establish a link between the
old couplings $\{\mu,v\}$ and the new ones $\{ v_{n,m} \}$.
By using \refE{AA5} repeatedly, we find
a relation among all the running couplings and reduce the whole
set to the independent ones.
This can be shown by noticing that all the running couplings
can be derived by knowing only $v_{n,0}$ for all $n$,
since the coupling $v_{n,m}$ is related  to $v_{n-1,m}$ and
$v_{n+1,m-2}$ by \refe{AA5}.
Thus, for any  positive integer $k$ we can draw a line
in the $n-m$ plane that starts at $m=0$ and $n=k$
and end at $m=2k$ and $n=0$.
If the running couplings, identified by the pair $(n,m)$, which lye
between the axes and this line are specified, then given $v_{k+1,0}$ one
can obtain all the couplings along the line $v_{n, -2n+2k}$ for $0 \le
n \le k+1$, thus extending the portion of the plane where the $v_{n,m}$
are known.
Proceeding in this way, all the running couplings can
be obtained if $v_{k,0}$ is known.
Thus $v_{k,0}$ can be chosen to be the independent
variable.
The usual action is recovered by choosing
$v_{1,0} = -2\alpha_l(\mu-v \alpha_l^2)$,
$v_{2,0} = -2(\mu-3v \alpha_l^2)$, and all others $v_{n,0}=0$
(for $\alpha_t=0$).
All other couplings can be found by the use of \refe{AA5}.

We consider now the local gauge invariance.
The invariance \refe{r60} holds for the original action.
We need to clarify the precise meaning of requiring that the
action written in terms of the new couplings $v_{n,m}$ is
invariant under $\refe{r60}$.
First, we face the problem of
defining a $\mu$ coupling, since there is no more trace of it
in the new action \refe{AA1}.
The implementation of the invariance \refe{AA5}
will thus involve a more complicated transformation of the fields, since
no simple coupling couples to $|\psi|^2$.
To find the proper transformation of the couplings, we need to consider the
transformation under an infinitesimal local gauge transformation
of the derivative term:
\beqa
    && \delta
    \left[
        i\, w_{lt} \int\!\!\! dx\, \psi_t(x) \partial_\tau \psi_l(x)
    \right]
    =
    \nonumber \\
    &&\qquad
    -i {w_{lt}\over 2} \int\!\!\! dx
    \left[
    \psi_l^2+\psi_t^2 + 2 \alpha_l \psi_l\right] \partial_\tau
    \delta\chi(\tau)
    \label{AA6}.
\eeqa
This implies the following transformation (with $\alpha_t=0$):
\beq
    \left\{
    \begin{array}{rcl}
    v_{2,0} &\rightarrow & v_{2,0} - i w_{lt}\,
    \partial_\tau \delta \chi(\tau)
    \\
    v_{0,2} &\rightarrow & v_{0,2} - i w_{lt}\,
    \partial_\tau \delta \chi(\tau)
    \\
    v_{1,0} &\rightarrow & v_{1,0} - i w_{lt} \alpha_l \,
    \partial_\tau \delta \chi(\tau)
    \,.
    \end{array}
    \right.
    \label{AA7}
\eeq
One can readily recognize in \refe{AA6} the field $|\psi|^2$ written
in terms of the new variables. An equivalent approach to define
the local $\tau$-gauge invariant is then to introduce a new external
source $\mu_{ext} \, w_{lt}/2$ that couples to
$|\psi|^2$ (which will be set to zero at
the end of the calculation).
In term of $\mu_{ext}$, gauge invariance is defined as usual by
$\mu_{ext} \rightarrow \mu_{ext}+\partial_\tau\, \chi(\tau)$.

We consider now the gauge transformations involving spatial variations.
In that case we have the following gradient terms in the action
\beq
    z_{tt}\int dx |\nabla \psi_t|^2 + z_{ll} \int dx |\nabla \psi_l|^2
    \label{AA8}
\eeq
where infinitesimal variation gives:
\beqa
    &&
    2\int\!\! dx
    \left\{
     z_{tt}\,\psi_l (\nabla \psi_t) \nabla \delta \chi(x)
    -z_{ll}\,\psi_t (\nabla \psi_l) \nabla \delta \chi(x)
    \right\}
    \nonumber \\
    &&
    \qquad+2\, (z_{tt}-z_{ll})
    \int\!\! dx
    (\nabla \psi_t)(\nabla \psi_l) \delta \chi(x)
    \label{AA9}.
\eeqa
The global gauge invariance of the action [with $\chi(x)=\chi$]
gives the condition $z_{ll}=z_{tt}$.
In this case, the resulting variation of the action reads:
\beq
    2\,z_{tt}
    \int dx
    \left\{
    (\psi_l \nabla \psi_t) -(\psi_t \nabla \psi_l)
    \right\}\nabla \chi(x)
    \label{AA10}
\eeq
that corresponds to the field  coupled linearly to $\bf A(x)$ in \refE{r1}.

In this way, we have clarified the precise meaning of {\em local} gauge
invariance in terms of the new couplings.
These results can be used, in particular, to determine the
composite-field insertions of Appendix \ref{App1loop}.

\section{Ward identities at one-loop level}
\label{App1loop}

In this Appendix, we compute explicitly the one-loop
contribution of relevant vertex functions. We can thus
analyze the validity of the WI through an explicit example.
In particular, the cancellations of the divergent contribution
in the WI will be demonstrated.

We begin with the analysis of the WI \refe{r77} when the coupling
$v_l$ is nonvanishing. We will start by considering all
running couplings present in the original action \refe{r1}.
The one-loop diagrams contributing to $\Gamma_{l}$
and $\Gamma_{tt}$ are:
\beqa
    \parbox{20mm}{\fmfframe(0,3)(0,0){
    \begin{fmfgraph*}(20,5)
    \fmfleft{i} \fmfright{o} \fmf{gll}{i,v1}
    \fmf{gtt}{v1,v1} \fmf{phantom}{v1,o}\fmfdot{v1}
    \end{fmfgraph*}}}
&=&
-{1\over 2}\, v_{ltt} \sum_q \Gc_{tt}(q) \,,
    \label{AB1}
\\
\parbox{20mm}{
\begin{fmfgraph*}(20,5)
 \fmfleft{i} \fmfright{o} \fmf{gll}{i,v1}
 \fmf{gll}{v1,v1} \fmf{phantom}{v1,o}\fmfdot{v1}
\end{fmfgraph*}}
&=&
-{1\over 2}\, v_{lll} \sum_q \Gc_{ll}(q) \,,
\label{AB2}
\\
\parbox{20mm}{\begin{fmfgraph*}(20,5)
 \fmfleft{i} \fmfright{o} \fmf{gtt}{i,v1,o}
 \fmf{gtt}{v1,v1} \fmfdot{v1}
\end{fmfgraph*}}
&=&
-{1\over 2}\, v_{tttt} \sum_q \Gc_{tt}(q)
\label{AB3} \,,
\\
\parbox{20mm}{\begin{fmfgraph*}(20,5)
 \fmfleft{i} \fmfright{o} \fmf{gtt}{i,v1,o}
 \fmf{gll}{v1,v1} \fmfdot{v1}
\end{fmfgraph*}}
&=&
-{1\over 2}\, v_{lltt} \sum_q \Gc_{ll}(q)
\label{AB4} \,,
\\
\parbox{20mm}{
\begin{fmfgraph*}(20,5)
 \fmfleft{i} \fmfright{o} \fmf{gtt}{i,v1} \fmf{gtt}{v2,o}
 \fmf{gtt,right,tension=.4}{v1,v2}
 \fmf{gll,right,tension=.4}{v2,v1} \fmfdot{v1,v2}
\end{fmfgraph*}}
&=&
-v_{ltt}^2 \sum_q \Gc_{ll}(k-q) \Gc_{tt}(q)
\label{AB5}
\\
\parbox{20mm}{
\begin{fmfgraph*}(20,5)
 \fmfleft{i} \fmfright{o} \fmf{gtt}{i,v1} \fmf{gtt}{v2,o}
 \fmf{glt,right,tension=.4}{v1,v2}
 \fmf{glt,right,tension=.4}{v2,v1} \fmfdot{v1,v2}
\end{fmfgraph*}}
&=&
-v_{ltt}^2 \sum_q \Gc_{tl}(k-q) \Gc_{lt}(q)
\,.
\label{AB6}
\eeqa
For non-vanishing $v_l$  the conditions on the bare
running couplings can be obtained from the WI \refe{r77},
\refe{r79}, and \refe{r69} (once Fourier transformed).
They give, respectively:
\beqa
    \psi_{lo} v_{tt} - v_l=0&& \label{con1} \,,
    \label{AB7}\\
    v_{ltt} \psi_{lo} +v_{tt}-v_{ll} = 0&& \label{con2}\,,
    \label{AB8}\\
    v_{lltt}\psi_{lo} + 2 v_{ltt}-v_{lll} = 0 && \label{con3} \,,
    \label{AB9}
\eeqa
while \refE{r88} is not modified (cf. \refE{r81}).
The validity of \refE{r77} at the one-loop order can be now
verified. Since $\Gamma_l$ is the sum of $v_{l}$,
\refe{AB1}, and \refe{AB2}, while $\Gamma_{tt}$ is the sum
of $v_{tt}$ and  \refe{AB3}-\refe{AB6}, we can readily evaluate
\refE{r77} for vanishing external momentum.
The lowest order is clearly satisfied by the bare values
$v_l=-2\alpha (\mu-v \alpha^2)$ and
$v_{tt}=-2(\mu-v \alpha^2)$. To verify that also the next
order (one-loop) is correct, we can use the following identity
\beq
    \Gc_{ll}(q)\Gc_{tt}(q) - \Gc_{lt}(q)\Gc_{tl}(q)
    =
    -{\Gc_{ll}(q)-\Gc_{tt}(q) \over v_{tt} - v_{ll}}\,,
    \label{AB12}
\eeq
holding for $u_{tt}=0$ and $z_{tt}=z_{ll}$.
Adding all the contributions and using \refe{AB12}, one can
verify that the WI hold {\em exactly}.

However, in obtaining a renormalizable theory,
all the irrelevant running couplings have been discarded
(for instance, $v_{lll}$ or $v_{lltt}$).
As anticipated in Section \ref{SecIIIa}, the WI are invalidated by
this fact but they continue to hold at the leading order
for $k\rightarrow 0$.
This can be verified by considering the contributions
to $\Gamma_l$ and $\Gamma_{tt}$ when the marginal running
couplings are set to zero.
In this case, the contributions are
\refe{AB1},\refe{AB3}, \refe{AB5}, and \refe{AB6},
and the WI {\em is not satisfied}.
We cannot test directly on this WI the cancellation of the leading
divergent terms, since $\Gamma_l$ and $\Gamma_{tt}(0)$ are finite.
One could, however, verify that the ultraviolet cutoff divergence is
still perfectly balanced in the WI; more instructive examples can
be obtained by looking at higher-order WI.

Let us consider accordingly the one-loop contributions to
$\Gamma_{ll}$, $\Gamma_{ltt}$, and $\Gamma_{tttt}$
due to the marginal couplings (from now on we set to
zero all the irrelevant running couplings).
On the same footing, the Green's functions are given
by the expression \refe{r57}
with $z_{ll}=0$, and $u_{tt}$ in general nonvanishing.
The mean-field condition is applied to the lowest-order values
for $v_{lt}$ and $v_{tt}$, which thus vanish.
Notice that, using the above conditions, \refe{AB12} becomes:
\beq
    \Gc_{tt}(q)\Gc_{ll}(q)-\Gc_{lt}(q)\Gc_{tl}(q) =
    -{\Gc_{tt}(q)\over v_{ll}} \,.
    \label{AB13}
\eeq
The one-loop diagram contributing to $\Gamma_{ll}(k)$ gives:
\beq
    \parbox{20mm}{
    \begin{fmfgraph*}(20,5)
     \fmfleft{i} \fmfright{o} \fmf{gll}{i,v1} \fmf{gll}{v2,o}
     \fmf{gtt,right,tension=.4}{v1,v2}
     \fmf{gtt,right,tension=.4}{v2,v1} \fmfdot{v1,v2}
    \end{fmfgraph*}}
    = - {v_{ltt}^2 \over 2} \sum_q \Gc_{tt}(k-q) \Gc_{tt}(q)
    \,.
    \label{AB14}
\eeq
We evaluate also the one-loop contributions to $\Gamma_{ltt}(k_1,k_2,k_3)$.
For simplicity, we write only the value of the diagrams for vanishing
external momenta, since the value for finite $k$ can be
reconstructed from the following expressions by inserting the correct
momenta in each internal Green's functions and adding the contribution of the
permutations of the external arguments.
We report below all the one-loop diagrams
for $\Gamma_{ltt}$:
\beqa
    \parbox{20mm}{
    \begin{fmfgraph*}(20,15)
     \fmfleft{i}\fmftop{o1}\fmfbottom{o2}
     \fmf{gll}{i,v}\fmf{gtt}{v2,o2}\fmf{gtt}{v1,o1}
     \fmf{glt,tension=.4}{v1,v}
     \fmf{glt,tension=.4}{v2,v}
     \fmf{gtt,tension=.1}{v1,v2} \fmfdot{v,v1,v2}
    \end{fmfgraph*}}
    &=&
      {v_{ltt}^3} \sum_q \Gc_{tl}(q)^2\Gc_{tt}(q)
    \,, \label{AB15}
    \\
2\ \ \parbox{20mm}{
\begin{fmfgraph*}(20,15)
 \fmfleft{i}\fmftop{o1}\fmfbottom{o2}
 \fmf{gll}{i,v}\fmf{gtt}{v2,o2}\fmf{gtt}{v1,o1}
 \fmf{glt,tension=.4}{v1,v}
 \fmf{gtt,tension=.4}{v2,v}
 \fmf{glt,tension=.1}{v2,v1} \fmfdot{v,v1,v2}
\end{fmfgraph*}
}
    &=&  {2\,v_{ltt}^3 } \sum_q \Gc_{tt}(q)\Gc_{tl}(q)
    \,, \label{AB16}
    \\
\parbox{20mm}{
\begin{fmfgraph*}(20,15)
 \fmfleft{i}\fmftop{o1}\fmfbottom{o2}
 \fmf{gll}{i,v}\fmf{gtt}{v2,o2}\fmf{gtt}{v1,o1}
 \fmf{gtt,tension=.4}{v1,v}
 \fmf{gtt,tension=.4}{v2,v}
 \fmf{gll,tension=.1}{v1,v2} \fmfdot{v,v1,v2}
\end{fmfgraph*}}
    &=&
     -{v_{ltt}^3} \sum_q \Gc_{tt}(q)^2\Gc_{ll}(q)
    \label{AB17}
    \\
    \parbox{20mm}{
    \begin{fmfgraph*}(20,5)
     \fmfleft{i} \fmfright{o1,o2} \fmf{gll}{i,v1} \fmf{gtt}{o2,v2,o1}
     \fmf{gtt,right,tension=.4}{v1,v2}
     \fmf{gtt,right,tension=.4}{v2,v1} \fmfdot{v1,v2}
    \end{fmfgraph*}}
    &=&
        - {v_{ltt}^2 v_{tttt} \over 2}
    \sum_q \Gc_{tt}(q)^2
    \,.     \label{AB18}
\eeqa
Only in diagram \refe{AB16} there are two different permutations.
With the same notation, we consider now the one-loop contributions to
$\Gamma_{tttt}$.
We write the following equation:
\beqa
    \Gamma_{tttt} &=&
    6 \parbox{15mm}{
    \begin{fmfgraph*}(15,15)
     \fmfleft{i1,i2}\fmftop{o1}\fmfbottom{o2}
     \fmf{gtt}{i1,v,i2}\fmf{gtt}{v2,o2}\fmf{gtt}{v1,o1}
     \fmf{glt,tension=.4}{v1,v}
     \fmf{glt,tension=.4}{v2,v}
     \fmf{gtt,tension=.1}{v1,v2} \fmfdot{v,v1,v2}
    \end{fmfgraph*}}
    +
12 \parbox{15mm}{
\begin{fmfgraph*}(15,15)
 \fmfleft{i1,i2}\fmftop{o1}\fmfbottom{o2}
 \fmf{gtt}{i1,v,i2}\fmf{gtt}{v2,o2}\fmf{gtt}{v1,o1}
 \fmf{glt,tension=.4}{v1,v}
 \fmf{gtt,tension=.4}{v2,v}
 \fmf{glt,tension=.1}{v2,v1} \fmfdot{v,v1,v2}
\end{fmfgraph*}}
+
6 \parbox{15mm}{
\begin{fmfgraph*}(15,15)
 \fmfleft{i1,i2}\fmftop{o1}\fmfbottom{o2}
 \fmf{gtt}{i1,v,i2}\fmf{gtt}{v2,o2}\fmf{gtt}{v1,o1}
 \fmf{gtt,tension=.4}{v1,v}
 \fmf{gtt,tension=.4}{v2,v}
 \fmf{gll,tension=.1}{v1,v2} \fmfdot{v,v1,v2}
\end{fmfgraph*}
}\nonumber
\\
&&
+
    3\parbox{20mm}{
    \fmfframe(0,4)(0,4){
    \begin{fmfgraph*}(20,5)
     \fmfleft{i1,i2} \fmfright{o1,o2} \fmf{gtt}{i1,v1,i2}
     \fmf{gtt}{o2,v2,o1}
     \fmf{gtt,right,tension=.4}{v1,v2}
     \fmf{gtt,right,tension=.4}{v2,v1} \fmfdot{v1,v2}
    \end{fmfgraph*}}}
+ S \,.
\label{AB19}
\eeqa
The numbers that multiply the diagrams indicate the non-equivalent
permutations of the external $k_i$. The contribution of these diagrams
is identical to the one given in \refe{AB15}-\refe{AB18},
apart from the number of permutations and an overall factor
$v_{tttt}/v_{ltt}$ to be multiplied.
Instead $S$ represents the sum of the following set of diagrams:
\beqa
6
\parbox{20mm}{
\begin{fmfgraph*}(20,15)
  \fmfleft{i1,i2} \fmfright{o3,o4}
  \fmf{gtt}{i1,v1}\fmf{gtt}{i2,v2}
  \fmf{gtt}{o3,v3}\fmf{plain}{o4,v4}
  \fmf{glt,tension=.3}{v1,v2,v4,v3,v1}
  \fmfdot{v1,v2,v4,v3}
\end{fmfgraph*}
}
 &=& -3\, v_{ltt}^4 \sum_q \Gc_{lt}(q)^4 \,,
    \label{AB20}
\\
24
\parbox{20mm}{
\begin{fmfgraph*}(20,15)
  \fmfleft{i1,i2} \fmfright{o3,o4}
  \fmf{gtt}{i1,v1}\fmf{gtt}{i2,v2}
  \fmf{gtt}{o3,v3}\fmf{plain}{o4,v4}
  \fmf{glt,tension=.3}{v1,v2,v4}
  \fmf{gll,tension=.3}{v4,v3}\fmf{gtt,tension=.3}{v3,v1}
  \fmfdot{v1,v2,v4,v3}
\end{fmfgraph*}}
 &=&
 -12\, v_{ltt}^4 \sum_q \Gc_{lt}(q)^2 \Gc_{ll}(q)\Gc_{tt}(q)
    \label{AB21} \,,
\\
12
\parbox{20mm}{
\begin{fmfgraph*}(20,15)
  \fmfleft{i1,i2} \fmfright{o3,o4}
  \fmf{gtt}{i1,v1}\fmf{gtt}{i2,v2}
  \fmf{gtt}{o3,v3}\fmf{plain}{o4,v4}
  \fmf{glt,tension=.3}{v1,v2}\fmf{gll,tension=.3}{v2,v4}
  \fmf{glt,tension=.3}{v3,v4}\fmf{gtt,tension=.3}{v3,v1}
  \fmfdot{v1,v2,v4,v3}
\end{fmfgraph*}
}
 &=&
 6\,v_{ltt}^4 \sum_q \Gc_{lt}(q)^2 \Gc_{ll}(q)\Gc_{tt}(q)
    \label{AB22} \,,
\\
6 \parbox{20mm}{
\begin{fmfgraph*}(20,15)
  \fmfleft{i1,i2} \fmfright{o3,o4}
  \fmf{gtt}{i1,v1}\fmf{gtt}{i2,v2}
  \fmf{gtt}{o3,v3}\fmf{plain}{o4,v4}
  \fmf{gtt,tension=.3}{v1,v2,v4}
  \fmf{gll,tension=.3}{v4,v3,v1}
  \fmfdot{v1,v2,v4,v3}
\end{fmfgraph*}}
 &=&
 -3\, v_{ltt}^4 \sum_q \Gc_{tt}(q)^2 \Gc_{ll}(q)^2
    \label{AB23} \,.
\eeqa
Adding all these contributions together and using \refe{AB13}, we finally
obtain for the $\Gamma$ functions at vanishing external $k$:
\beqa
    \Gamma_{ltt} &=&
    - {1\over 2} { v_{ltt}^2 \over \psi_{lo}^2}
    \sum_q \Gc_{tt}(q)^2
    \,,
    \label{AB23b}
    \\
    \Gamma_{tttt} &=&
    -{3\over 2} {v_{ltt}^2 \over \psi_{lo}}
    \sum_q \Gc_{tt}(q)^2
    \label{AB24} \,.
\eeqa
The above expressions are divergent when $d \leq 3$.
An IR cutoff should thus be introduced, with the vertex functions
diverging when this cutoff goes to zero.
It is readily verified that the leading diverging contributions
of each vertex function simplify exactly when \refe{AB23b},
\refe{AB24}, \refe{AB14}, and \refe{AB3} are inserted into
\refe{r79} and \refe{r81}.
Note, however, that there exist finite terms
(specifically, the contribution from $\Gamma_{tt}$)
 that do not vanish, so that the WI is not
exactly satisfied.
This fact had to be expected and does not spoil the RG
identity among the running couplings, since within the scheme of the
dimensional renormalization they are valid for the diverging
contributions.
The PT is thus preserving the gauge invariance at the
leading order in the IR divergences.

We consider now \refE{r76} as the simplest example of
cancellation  of the diverging coefficients of
the external frequency $k_o$.
To verify this, we need to evaluate $\Gamma_{lt}$ and $\Gamma_{l;0}$
for finite $k_o$. The one-loop contribution is:
\beqa
    \Gamma_{lt}(k)
    &=&
    \parbox{20mm}{
    \begin{fmfgraph*}(20,5)
     \fmfleft{i} \fmfright{o} \fmf{gll}{i,v1} \fmf{gtt}{v2,o}
     \fmf{gtt,right,tension=.4}{v1,v2}
     \fmf{glt,right,tension=.4}{v2,v1} \fmfdot{v1,v2}
    \end{fmfgraph*}}
    =
    -v_{ltt}^2 \sum_q \Gc_{tt}(k-q)\Gc_{lt}(q) \,,
    \nonumber \\
    \label{AB25}
    \\
    \Gamma_{l;0}(k)
    &=&
    \parbox{20mm}{
    \begin{fmfgraph*}(20,5)
     \fmfleft{i} \fmfright{o} \fmf{gll}{i,v1} \fmf{photon}{v2,o}
     \fmf{gtt,right,tension=.4}{v1,v2}
     \fmf{gtt,right,tension=.4}{v2,v1} \fmfdot{v1,v2}
    \end{fmfgraph*}}
    =
    - {1\over2} v_{ltt} w_{lt} \sum_q \Gc_{tt}(k-q) \Gc_{tt}(q) \,.
    \nonumber \\
    \label{AB26}
\eeqa
When we evaluate \refE{r76} for $k_i=0$ and finite $k_0$, we
obtain (we have used $\Gc_{lt}(k)=\Gc_{tt}(k) k_0 w_{lt}/v_{ll}$ from
\refe{r57}):
\beq
    -{1\over 2} v_{ltt} w_{lt} \sum_q \Gc_{tt}(k-q)\Gc_{tt}(q)
    (q_o-k_o/2)\,.
    \label{AB27}
\eeq
In order to extract the $k_0$ coefficient,
let us perform the change of variables $q=q'+k/2$.
Since an ultraviolet cutoff $\Lambda$ is assumed, the above change
of variable should change the limits of integrations. We neglect this
fact, since the difference is of higher order in $q_0/\Lambda$, and we
are interested in the $q_0\rightarrow 0 $ limit.
With the above change of variable, we readibly obtain that the leading
contribution of \refe{AB27} vanishes.

It is not difficult to evaluate also $\Gamma_{l;i}(k_o,\kv)$ and
insert the result into \refE{r76}. We can evaluate then all
$\Gamma$ functions at finite values of $k_o$ and $k_i$.
The above results remain valid for the first part of the WI, since no
use was made of the fact that $\kv_i$ was vanishing.
However, the contribution of the vector part of the WI is sub-leading
with respect the one calculated in \refe{AB27}.
To check the vanishing of the vector part, one should thus perform
the above calculation by keeping also the next-to-leading order.

\section{one-loop RG equations}
\label{AppRG1loop}

In the text we have shown that a single
coupling is actually scaling independently.
In this Appendix, we obtain the one-loop RG equations for the four
marginal couplings entering the propagators ($v_{ll}$, $w_{lt}$,
$u_{tt}$, and $z_{tt}$) and verify that the exact constraints found in
Section \ref{addconst} are fulfilled.
We need to calculate explicitly the one loop contributions to
$\Gamma_{ll}$ [depicted in \refe{AB14}],  $\Gamma_{lt}$ [depicted in
\refe{AB5}], and  $\Gamma_{tt}$ [depicted in \refe{AB5}, and
\refe{AB6}].
According to the general procedure established in Section \ref{SecIIB}
no correction to $\alpha$ need to be considered.
The integrals entering these expressions can be written in terms of
the following integrals (when the expression for the propagators
in the small-momentum limit is considered):
\beq
   I_{n,m}(k)
   = \int {d^{d+1} q \over (2\pi)^{d+1}}
   {q_o^n (\qv^2)^{m/2} \over (k+q)^2 q^2}
   \label{AC1}
\,.
\eeq
We thus have at the one loop order:
\beqa
   \Gamma_{ll}(k) &=& v_{ll} -
   {v_{ll}^2 \over 2 \psi_{lo}^2  z_{tt}^2 }  I_{0,0}(k)
   \label{AC2}
   \\
   \Gamma_{lt}(k) &=& w_{lt} k_0
   +{w_{lt} v_{ll}  \over \psi_{lo}^2  z_{tt}^2 }  I_{1,0}(k)
   \label{AC3}
   \\
   \Gamma_{tt}(k) &=& u_{tt} k_0^2 + z_{tt} \kv^2
   -{ 1 \over  \psi_{lo}^2  z_{tt}^2}
   \left[
   (u_{tt} v_{ll}+w_{lt}^2)  I_{2,0}(k)
   \right.
   \nonumber \\
   &&
   \left.
   + z_{tt} v_{ll} I_{0,2}(k)
   + w_{lt}^2  k_0 I_{1,0}(k)
   \right] \,.
   \label{AC4}
\eeqa
The $I_{n,m}$ integrals are performed by standard procedures:\cite{Amit}
\beqa
  I_f(k)
   &=&
   {\Gamma(2) \over \Gamma(1)^2}
   \int {d^{d+1} q \over (2\pi)^{d+1}}
   \int_0^1 \!\!\! dx
   {f(q) \over [x(k+q)^2 +(1-x)q^2]^2}
   \nonumber
   \\
   &=&
   {1\over 2}
   \int_0^1 \!\!\! dx
   \int {d^{d+1} q \over (2\pi)^{d+1}}
   {
   f(q-kx)+f(-q-kx)
   \over
   [q^2 +k^2 x(1-x)]^2
   }
   \nonumber
\eeqa
where $f(q)=q_o^n \qv^{m}$.
One can now use the identity
\beq
\int_0^{+\infty}
\!\!\!
{x^{d} dx \over (1+x^2)^\alpha} =
{
\Gamma((d+1)/2)) \Gamma(\alpha-(d+1)/2) \over 2 \Gamma(\alpha)
}
\label{AC5}
\eeq
to obtain the analytic expression to be continued
to complex values of the variable $d$.
As a matter of fact, the integral \refe{AC5}
converges only for $d-2\alpha>0$, but its analytic continuation
given by the Euler $\Gamma$-functions on the right-hand side of
\refe{AC5} is defined for any value of the complex variable $d$, except
for a discrete set of points where simple poles are present.
Dimensional regularization proceeds by subtracting these
poles that in our case are in $1/(3-d)=1/\epsilon$.
We thus expand the integrals in $\epsilon$ keeping only
the leading order.
This gives the following expressions for
$\Gamma_{ij}$ (we have absorbed the phase-space factor $1/(8\pi^2)$ in
the definition of the running couplings
$v_{ll}\rightarrow v_{ll}/8\pi^2$
and $w_{lt}\rightarrow w_{lt}/\sqrt{8\pi^2}$):
\beqa
   \Gamma_{ll}(k) &=& v_{ll} -
   {v_{ll}^2 \over 2 \psi_{lo}^2  z_{tt}^2 }
   {(k^2) \over \epsilon}^{\epsilon/2} {c\over c_o}
   \nonumber
   \\
   \Gamma_{lt}(k) &=& w_{lt} k_0
   -{w_{lt} v_{ll}  \over 2 \psi_{lo}^2  z_{tt}^2 } k_0
   {(k^2) \over \epsilon}^{\epsilon/2}
   \left({c\over c_o}\right)^2
   \nonumber
   \\
   \Gamma_{tt}(k) &=& u_{tt} k_0^2 + z_{tt} \kv^2
   +
   {
   w_{lt}^2 k^2_0
   \over
   2 \psi_{lo}^2  z_{tt}^2
   }
   {(k^2) \over \epsilon}^{\epsilon/2}
   \left({c\over c_o}\right)^3
   \nonumber
\eeqa
where $c^2$ indicates the combination of running couplings
forming the square of the sound velocity [cf. \refE{s6}].
As a first result, we note that no divergence appears in
the $\kv^2$ dependence.
This implies that $z_{tt}$ does not scale, as
demonstrated using the WI in Section \ref{localWI}.
The RG equations obtained by minimal subtraction
for the remaining three couplings read:
\beqa
   \rho {dv_{ll} \over d \rho}
   &=&
   {v_{ll}^2(\rho) \over 2 \psi_{lo}^2(\rho) z_{tt}^2}
   {c(\rho)\over c_o}
   \\
   \rho {dw_{lt} \over d \rho}
   &=&
   {v_{ll}(\rho) w_{lt}(\rho) \over 2 \psi_{lo}^2(\rho) z_{tt}^2}
   \left({c(\rho)\over c_o}\right)^2
   \\
   \rho {du_{tt} \over d \rho}
   &=&
   -{w_{lt}^2(\rho) \over 2 \psi_{lo}^2(\rho) z_{tt}^2}
   \left({c(\rho)\over c_o}\right)^3
\eeqa
where $\psi_{lo}(\rho)=\psi_{lo}(1) \rho^{\epsilon/2}$ due to its
bare dimension.
It is now possible to verify that the RG equations admit the
invariants $ {w_{lt}(\rho)/v_{ll}(\rho)} $ and $c^2(\rho)$, by
simply performing the derivative with respect to $\rho$ and
substituting the RG equations.
This shows  at the one-loop level  the validity of the identities proved
at all orders in Section \ref{addconst}.
These identities can be used to eliminate two out of the three running
couplings. For the last one ($v_{ll}$, for instance), we need to solve
the RG equation explicitly.
The solution is given and discussed in Section \ref{oneloop}.

\widetext

\end{fmffile}

\begin{references}
\bibitem{mufixed}

The noninteracting boson system, on the contrary,
with fixed chemical potential is not well defined
(as at $T=0$ and $\mu<0$ the system contains no particle).
As a consequence, the perturbative expansion is not well defined.
On the other hand, the organization of classes of diagrams
on the basis of the IR divergences -as we do- seems to
be a well defined procedure since it bypasses the expansion
in the bare potential.
\bibitem{phasefree}  One can show by power counting
that in the variables phase and modulus  all running couplings
are irrelevant. This is equivalent to stating that no IR divergences
appear in the theory.
\end{references}
\end{document}